# Blockchain-Based Privacy-Preserving Data Management

Michael Mireku Kwakye*

*Department of Computer Science and Informatics, Fort Hays State University, Hays, Kansas 67601, USA*

**Abstract:** Privacy-preservation policies are guidelines formulated to protect data provider's private data. Previous privacy-preservation methodologies have addressed privacy in which data are permanently stored in repositories and disconnected from changing data provider privacy preferences. This occurrence becomes evident as data moves to another data repository. Hence, the need for data providers to control and flexibly update their existing privacy preferences due to changing data usage continues to remain a problem. This paper proposes a blockchain-based methodology for preserving data provider's private and sensitive data. The research proposes to tightly couple data provider's private attribute data element to privacy preferences and data accessor data element into a privacy tuple. The implementation presents a framework of tightly-coupled relational database and blockchains. This delivers secure, tamper-resistant, and query-efficient platform for data management and query processing. The evaluation analysis from the implementation validates efficient query processing of privacy-aware queries on the privacy infrastructure. Moreover, a comparative analysis to other methodology approaches indicates that our approach offers efficient and better technique for data provider's private data protection and identity preservation.

**Keywords:** Privacy model, Privacy infrastructure, Blockchains, Privacy-preserving data management.

## 1. INTRODUCTION

Privacy-preservation in data repositories is a fundamental and necessary requirement in the processing of personalized, private data across varied data management areas. The concept of preserving privacy in data stores outlines methodologies and implementations that help protect data provider's private and sensitive; during data storage, management, and query processing in the data stores. Hence, privacy-preservation methods offer data collectors and/or accessors the platfrom to process (and draw data utility) from data provider's private and transaction data without breaching privacy and personal identification. In practical terms, the discussion and applications of privacy concepts has significant influences in different domains, such as, legal frameworks, healthcare management, and government data platforms, amongst others. The emergence of multiple diverse data collection points across different application domains allows the processing of large data volumes on data providers. Most especially, associated transactions from collected data provider's private, sensitive data through point-of-contacts user interface platforms opens up important requirements for storage and management of these data.

Consider the application domain of a healthcare system where patient healthcare records are collected and maintained for healthcare service delivery. The primary source of data is from the patient and so we classify the patient as the data provider. Different forms of data are collected from the data providers. These collected data may be hypothetically categorized as less-sensitive, medium-sensitive, and highly-sensitive. The categorization of these collected data defines an efficient specification of privacy preferences for related data of the same sensitivity level. The varied data forms enable efficient methods for storage, management, and disclosure of related patient private, sensitive data items. It is noted that based on the information from the application domain and case study requirement analysis, we adopt these attribute data categorizations. For example, data items, such as, *health_insurance_carrier* and *consent_witness_name* could be characterized as less-sensitive data, whiles *patient_last_name* and *postal_code* could be categorized as medium-sensitive data. Data items, such as, *date_of_birth* and *personal_health_number* could be categorized as highly-sensitive data. Additionally, collected data in healthcare domain may be categorized as biographical information (*i.e.*, *first_name*, *last_name*, *date_of_birth*, and *gender*, amongst others), demographic information (*i.e.*, *home_address*, *postal_code*, *province*), patient consent information (*i.e.*, *witness_biographic_data*, *witness_address*, *dependent_data*, and *allergies*, amongst others), and health insurance information (*i.e.*, *carrier_name*, *policy_number*, and *group_number*).

*Address correspondence to this author at the Department of Computer Science and Informatics, Fort Hays State University, Hays, Kansas 67601, USA; E-mail: m_mirekukwakye@fhsu.edu

The healthcare system presents different categories of data collectors or service providers (*i.e.*, clinicians, lab analysts, *etc.*) who collect disparate data items at separate stages of healthcare delivery. Moreover, we identify data accessors who are designated as individuals (or groups) that request data access and process the collected data. In some instances, data collectors become data accessors due to their work needs and responsibilities.

Processing massive data volumes demands enhanced operational efficiency, performance, and optimal service delivery for both data and service providers (and third-party data accessors). Within the healthcare domain, where private data management is vital, issues of information security and trust between parties involved in service delivery must be adequately addressed. This motivates the need to secure data access and authenticate users on data management systems. Conversely, the need to protect data provider's data privacy arises with the processing of such private, sensitive data provider healthcare data, such as, cancer-related data [32]. Hence, demand for efficient data privacy merits better privacy-preservation policies and methodologies to address critical service delivery on these data processing platforms.

Previous privacy-preservation methodologies address privacy for data permanently stored in repositories -sometimes called "data-at-rest." Data provider privacy preferences are rarely considered (especially, as the data moves to another data repository or third-party accessor) [2]. We address the problem statement from two main areas. First, with the persistence of these static data in data repositories, data providers or data collectors infrequently or minimally respond to or are alerted to changes in data provider privacy preferences. The addition of new data records or expectation of new privacy preferences on new and/or existing data poses a major challenge in handling data provider privacy preferences [6]. Second, inability of data providers to flexibly update or change their existing privacy preferences when desired, or when data usage changes. As a result, the inability to track and/or control privacy violations by service providers or third-party accessors continues to remain a problem.

Managing and protecting private information in data repositories has generated much research interest since the seminal work on *hippocratic databases* [22]. Research on *contextual integrity* (CI) and purpose-based privacy policies in data provider privacy preferences provide results that seem to suggest improved methodologies and models in data privacy management [13, 14, 15, 21, 33]. Despite the results obtained from contextual integrity and purpose-based privacy, the problem of a practical implementation to protect data provider's privacy preferences for various application domains remains a challenge. Moreover, a privacy-preservation approach for managing data provider's private, sensitive data that adopts an immutable, tamper-resistant data platform to process data provider's privacy preferences on large data volumes is yet to be devised. The approaches studied so far do not adopt an ontology framework to model data provider preferences and privacy semantics in their data protection policies. Moreover, current privacy ontologies provide insufficient semantics to support the complex nature of data provider privacy preferences. The ontologies lack expressive language and semantics to describe data provider preferences [12, 38]. These prior approaches based on contextual integrity and purpose-based privacy are not easily applied to real-world scenarios.

This paper addresses a blockchain-based privacy-preservating platform for data management (regarding storage and query processing) for data provider's private, sensitive data. The technical contribution involves an approach that offers control over the usage of sensitive private data by data providers, data collectors, and/or permitted third-party data accessors. The novel concept adopts a formal contextualized privacy ontology; which is foundational to the methodology. Additionally, the approach adopts to tightly couple and encapsulate attribute data, data provider privacy preferences, and data accessor's profile. This is implemented using object-oriented methods. The technological framework of the approach involves coupling both blockchains and relational database platforms.

### 1.1. Our Contributions

The technical contributions are summarized as follows:

- We introduce a privacy framework where data providers have ultimate access and control of their privacy preferences, and every change request is validated by both data providers and data collectors (or service providers). This is achieved through an integrated privacy tuple generated from the tight-coupling of data elements;

- We introduce an efficient procedure to tightly couple attribute data, data provider privacy preferences, and data accessor data elements into a privacy tuple. The tightly coupled privacy tuple is transmitted through an encrypted application transmission medium and stored in an immutable data platform, such as blockchain;

- We introduce an efficient coupling between relational database management system and private blockchains platform; as part of an encrypted, secured data processing platform for data accessors. We also discuss the implementation details of the system architecture. We establish that the blockchain is feasible to use as an authentication platform for all forms of data access and changes to data provider's privacy preferences;

- We introduce a privacy model framework that uses a formulated, dynamic contextualized privacy ontology as a foundational model. We describe the ontology's key characteristics of *purpose*, context, data provider privacy preferences, user roles, contextual norms, transmission medium, and other privacy-related factors that may be relevant in a data provider's privacy policy.

This paper has been significantly extended from previous conference proceedings [1]; which outlines a much summarized publication of the overall methodology approach. The work presented here represents an comprehensive discussion on related background research, elaborate discussion of the methodology, detailed implementation procedures, and discusses additional in-depth evaluation analysis – with comparative analysis to other related privacy-preserving methodology approaches. In addition, this paper discusses the overall system security assessment, the security/adversarial model, the privacy infrastructure, and the methodology merits in comparison to other approaches.

The remainder of the paper is organized as follows. In Section 2, we discuss background studies on data privacy models, privacy ontologies, blockchains, and relational databases. Section 3 discusses key contributions and methodology overview for a privacy model framework. Section 4 discusses the implementation of the proposed research methodology and Section 5 discusses the evaluation and analysis of the results of the implementation procedures performed. Section 6 discusses a comparison of our methodology approach to other methodologies. Here, we assess efficient and better measures for data provider's private data protection and identity preservation in relation to other approaches. Section 7 summarizes our contributions and discusses open issues and future work.

## 2. RELATED BACKGROUND RESEARCH

Privacy preservation approaches provide useful outcomes and/or results in policies, techniques, algorithms, and useful knowledge in securing data provider's private data in data repositories. We discuss the related background research work regarding privacy models and methodologies, privacy ontologies and contextual privacy, blockchain-based privacy data management.

### 2.1. Privacy-Preserving Models and Methodologies

Several data privacy-preservation approaches have been addressed and proposed by researchers in the data privacy domain. The development of privacy-preserving anonymization methods prevents disclosure of private, sensitive information of persons in published data. One early study on privacy-preserving anonymization methods is provided by Sweeney's *k-anonymity* [3]. There has been a succession of other approaches that seek to address the weaknesses in earlier methods. These models and methodologies are presented in different anonymization techniques (such as, generalization and suppression [3, 4], randomization/data perturbation [5], and data swapping [11]).

Privacy-preserving data mining methodologies are techniques and algorithms that are employed to protect private, sensitive data while drawing relevant knowledge from underlying data repositories. Aggarwal and Yu [6] provide a foundational study on privacy-preserving data mining. The authors propose an approach that does not require recent problem-specific algorithms, but maps and condenses groups of original data sets into new anonymized set. The approach uses statistical information to create anonymized data which has similar statistical characteristics. Subsequent investigations of privacy-preserving data mining discuss critical areas in handling broad data distributions. Additionally, other studies analyse the trade-off between sharing information and privacy-preservation, and generate valuable methodologies and results [8, 9, 10].

The study of better privacy-preserving and privacy-aware database models have garnered interest in the research community since the seminal work on *Hippocratic databases* [22]. Agrawal *et al.* [22] outline and address several principles for protecting data stored in database management systems. Moreover, the authors propose an architectural design framework (in terms of privacy meta-data schema, privacy policy schema, privacy authorization schema, *etc.*). This design models a privacy-aware data repository where measures for data collection, query processing, and data retention, amongst others, are outlined. Further studies by Barker *et al.* [23] seek to outline a privacy taxonomy for the collection, sharing, and disclosure of private data stored in repositories. The authors develop a privacy taxonomy to define parameters for a privacy policy while protecting the data provider's personalized information. Barker's [24] conceptual approach to privacy protection addresses the need to identify trade-off between maximizing data utility and optimum privacy in data repositories. His study proposes a conceptual approach where data provider approves or consents to privacy policies or changes in existing policies with the service provider.

## 2.2. Privacy Ontologies, Contextual Privacy, and Purpose-Based Privacy

Privacy ontologies form the structure and model for most contextbased privacy policy formulation. Ontologies formalize diverse preferences, unique attributes, and distinct rules for managing, processing, and modelling privacy policies. Several research efforts have investigated ontologies, and these studies define and describe how ontologies are used in modelling real-world scenarios and application domains. Bawany and Shaikh [12] model a data privacy ontology for various ubiquitous computing platforms. Their research work defines ontology semantic constructs (of *data*, *producer*, and *consumer*) and predicates for privacy preservation of data throughout its life cycle. Other studies propose ontology modelling to protect private, sensitive data for various application domains, such as, healthcare delivery, supply chain services, and financial services, amongst others. Some studies connect domain-specific ontology modelling to associated processing platforms, such as, mobile devices, and Point-of-Contact (PoC) user interfaces [34]. Accordingly, the study of context-aware and rule-based ontologies offer many results that address such unique scenarios. Moreover, the outputs of context-aware and rule-based ontologies are relevant in defining the semantics and predicates of privacy ontology modelling [35, 36, 37].

The research paradigm of *contextual integrity* (formalized by Nissenbaum *et al.* [13, 14]), focus on context-specific viewpoints, such as, *role*, *temporal norms*, *value*, and *transmission principle*, among others; to ensure integrity of information flow. These viewpoints provide further insights and perspectives to modelling of privacy ontologies and frameworks. The proposition by Nissenbaum seek to outline different context-based privacy norms adopted in virtual communities and data processing paradigms. Subsequently, Jafari *et al.* [15, 21] investigate purpose-based privacy policies and modelling to formulate workflow models which are used as frameworks for privacy policies. Their studies address different insights and motivations behind *purposes* for application domains. Moreover, the authors propose a formal definition for purpose (being a condition of reason among inter-related actions) and its constraints. The authors develop a framework model to express and enforce such purposes in privacy policies. Their formal framework further discusses the properties of purpose and shows how purpose-based constraints can be connected to more general access control policies. Finally, the authors specifically address enforcement of purpose within the framework of purpose-based privacy using workflow-based information systems. Other contributions in using purpose lattices and application of purpose in graph modelling are outlined by Kumar and Shyamasundar [18], Tian *et al.* [19], and van Staden and Olivier [20].

## 2.3. Blockchains and Data Privacy

Blockchains offer secured, tamper-resistant, and immutable framework for data processing. These platforms provide a better "database" framework for processing private, sensitive data provider's data. A number of studies have investigated a background framework and design of blockchains data platform. Romano and Schmid [7] discuss the adoption of the blockchain technology platform in diverse data processing systems. In their study, the authors discuss the current and potential application of blockchains for different domains. The authors discuss smart property management, where private information, such as, copyrights and patents of data owners are protected. Zyskind *et al.* [16] and Halpin and Piekarska [17] independently investigate aspects of blockchains security, encryption, and extensions to protecting personal data within the framework of decentralized processing. In these works, the research identifies ways in which users (data providers) can own and control their data whiles having service providers offering them personalized services [16]. The research addresses how to protect private data using blockhains as an access-control moderator with an off-blockchain storage solution.

These studies outline unique benefits of offering data verifiability, identity, immutability, encryption, and decentralization [25]. Notable research combining the merits of both blockchains and big data database platforms is addressed by McConaghy *et al.* [26]. In their studies, the authors simulate coupling distributed, decentralized blockchains and big data database systems. Some prior research work that use blockchain platform to protect data provider's data and preserve privacy and identities of these data providers have been addressed in the literature. These approaches adopt varied methodologies in their privacy modelling and architecture.

Prominent research works by Daidone *et al.* [27], Fernandez *et al.* [28], and Griggs *et al.* [29] present important methodologies and outcomes with regard to the adoption of blockchains to facilitate preservation of data provider's private data. Daidone *et al.* [27] propose a blockchain-based privacy enforcement architecture where users can define how their data are collected and managed; and ascertain how these data are used without relying on a centralized manager. Their methodology adopts a blockchain to perform privacy preferences' compliance checks in a decentralized fashion on IoT devices; whiles ensuring the correctness of the process through smart contracts. Fernandez *et al.* [28] propose a cloud-IoT architecture, called *Data Bank*, that aims at protecting users' sensitive data by allowing them to control which kind of data is transmitted by their devices. Their architecture consists of several layers of IoT objects, web and mobile applications, and a cloud layer for enforcement and access control. Griggs *et al.* [29] propose a healthcare blockchain system for secure automated remote patient monitoring. In their approach, the research outlines using blockchain-based smart contracts to facilitate secure management of medical sensors. Their architecture uses a system where the sensors communicate with a smart device that calls smart contracts, and writes records of all events on the blockchain.

### 2.4. Research Gaps and Summary

Previous research has found some methods of protecting data provider's private and sensitive information. However, a better blockchain-based approach to efficiently manage and control data access, availability, and usage of data provider's private, sensitive data stored in relational databases is still a problem. This is because of: (a) the absence of an expressive privacy model and/or ontology model, (b) lack of an adaptable privacy infrastructure and architectural platform, (c) lack of adoption of an expressive and robust private blockchain platform, and (d) minimal degree of data protection provided by data collectors (*i.e.*, service providers). Moreover, the ability to ensure data integrity and data monitoring to changes to data provider's privacy preferences for data use by data collectors/accessors (and consenting third-party accessors) continue to remain a challenge. A formulated methodology in which there is tight-coupling of private, sensitive data, with contextualized data provider privacy preferences, and data accessors' profile has not been addressed yet. Furthermore, a technology that couples blockchains and relational database management system is yet to be devised. This coupling approach will enable the storage of data provider's privacy preferences on their private, sensitive information in a completely tamper-resistant, immutable, and untrusted platform.

## 3. PROPOSED PRIVACY MODEL AND INFRASTRUCTURE

We describe and discuss details of our proposed privacy model and infrastructure. The model offers an effective integration of system components to provide a platform where data provider's information and transaction data are collected, processed, and managed by data collectors and/or accessors.

### 3.1. System Methodology Overview

The detailed outline of the methodology is as follows: (a) formulation of a formal contextualized and usable privacy ontology; (b) tight-coupling of data elements (attribute data, data accessor profile, and privacy preferences); (c) tight-coupling of decentralized blockchains and a relational database; (d) query processing required for proposed methodology; and (e) encryption approaches on system architecture of (blockchains and relational database). Each of these system components and methodologies work together to provide privacy-preservation in a privacy-aware data platform.

The overall procedure for the methodology involves the following steps: data elements collection, binding data elements into a privacy tuple, applying a hash function to the privacy tuple, storing the hashed privacy tuple on the blockchains, storing data provider data and transactional data in the relational database, and executing queries against privacy-aware stored data with reference to privacy tuples. In terms of data elements collection, the attribute data element, data provider's privacy preferences data element, and data accessor profile data element are collected from system user interfaces. First, privacy-related personal data values are collected from the data provider through point-of-contact user interfaces. Devices designated as point-of-contact interfaces for data collection are portable mobile devices and personal computing platforms. The privacy-related data values may come from biographical, demographic, financial, and healthcare information data. Additionally, other application domain transaction data are collected.

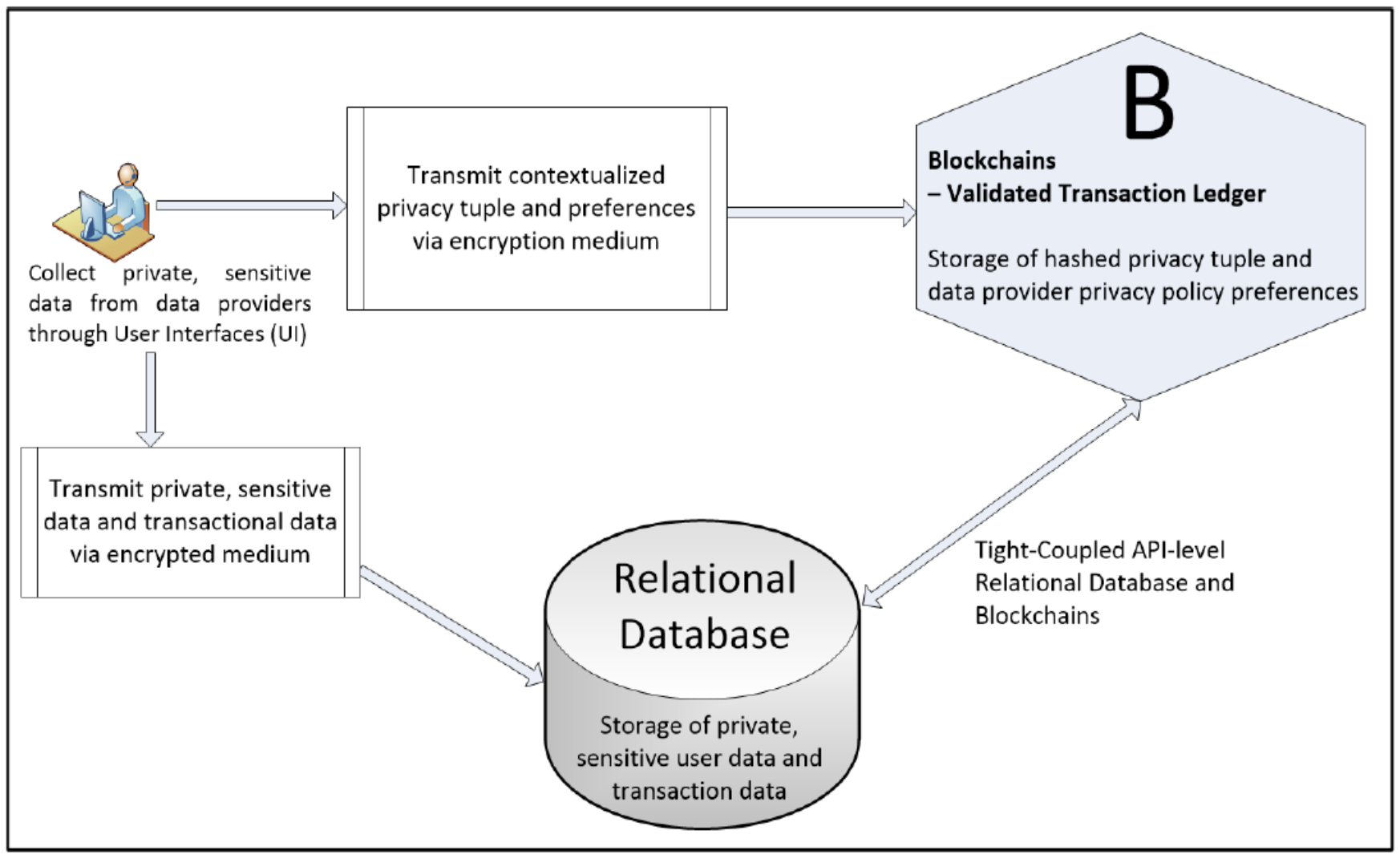

**Figure 1:** System Methodology Overview and Architecture.

The data provider agrees to the privacy policy preferences needed to access data utility services by the data collector (or service provider). Once the data elements are collected, attribute meta-data (from the private, sensitive data) is extracted and bound together with data provider's privacy preferences data (from the privacy policy) and data accessor profiling data, to form a *privacy tuple*. The privacy tuple is hashed, saved, and permanently stored on the blockchain platform, while the collected data provider private data and transactional data are stored in the relational database. The transmission of hashed privacy tuples and raw data provider private data to the blockchain platform and relational database, respectively, are facilitated through encrypted data communication channels that exist between the system components: relational database, blockchain platform, and user interfaces. The general architecture of the blockchain platform (which involves the transaction ledgers, data streams, and data items) are configured in a way that the data items are directly accessible to every (authenticated) user who has secured access to query the blockchain node. Hence, the adoption of hashing methodology on the privacy tuples before they are stored on the blockchain provides another layer of data protection to the privacy tuples.

During query processing, hashed privacy tuples are retrieved from the blockchains. These hashed privacy tuples are verified against the privacy tuples stored in the relational database (to check for data integrity). In cases where there is data inconsistency in the verification process of the hashed privacy tuples from the blockchains to the privacy tuples retrieved from the relational database, the query is aborted or otherwise the overall query retrieval process is allowed to proceed. Based on the data value composition in the privacy tuples, all privacy preferences (which are outlined in the privacy policy and) related to the attribute data to be queried are enforced. The query parsing process is subsequently allowed, and the required privacy-aware data records are then retrieved and provided to the requesting data accessor. Figure **1** provides a general overview of the methodology. We discuss the detailed overview of the query processing in the methodology in Section 3.5.

### 3.2. Formal Contextualized Privacy Ontology

The overall methodology approach formulates a contextualized privacy ontology with unique semantics and properties for privacy preservation. This privacy ontology serves as a basis to formalize the privacy model for data processing. Moreover, the formal contextualized privacy ontology becomes the foundation for the design of all entities and predicates in the data privacy-preservation process. This is necessary as the privacy ontology offers a modelling framework for procedural activities and semantics for private, sensitive data and querying processing by data collectors and third-party data accessors.

In the formulated privacy ontology, entity classes, sub-classes, their respective attributes, and associated properties are classified to handle prevalent privacy-related scenarios in privacy-preserving data management. One important aspect that must be considered in a formal contextualized privacy ontology is the need to address semantics in the ontology. To this end, the formulated privacy ontology model details and defines language structure, semantic knowledge, and formalisms of the entities. These modelling semantics handle privacy and data protection authorizations of related *contextual privacy norms* and *perspectives* (of *when*, *who*, *why*, *where*, and *how*). Moreover, the formal contextualized privacy ontology addresses sensitive data provider data usage and authorizations applicable to various data processing within the service delivery.

We designate the primary (*root*) entity item as *AttributeDataValue* in the formulated privacy ontology. This is because each attribute data has a unique privacy policy for data storage and retrieval. The *AttributeDataValue* entity has three sub-entity classes; namely, *DataProvider*, *DataCollector*, and *DataPrivacy Policy*. Figure **2** illustrates a summarized description of the formal contextualized ontology. The *DataProvider* sub-entity describes all information regarding the data provider, (for example, patient requesting healthcare service). This sub-entity has a predicate class, *DataProviderType*, which details data on the various forms of persons who serve as the primary source of data. The instantiated values (within healthcare application domain) for this predicate class are *Patient* and *PatientLegalRepresentative*; where *Patient* identifies with the patient who requests for medical service and offers the necessary data for service delivery. The *PatientLegalRepresentative* is a legal representative for the patient, in cases where the patient is a minor or does not have the capacity to offer requested data. This individual may be a parent, family representative, or a close friend.

The *DataCollector* sub-entity describes the individuals who collect and store data on data subjects (*i.e.*, data providers). This sub-entity also has a predicate class as *DataCollectorType*, which has instantiated values as *ClinicalNurse*, *Laboratory Analyst*, and *ClinicalPhysician* in a healthcare system.

The *DataPrivacyPolicy* sub-entity is the most expressive sub-class of the *AttributeDataValue* class. It describes all aspects of the privacy policy on each attribute data value; which has been fully authorized by the data provider and validated by the data collector. We model this entity class based on prior research work of Barker *et al.* [23] on privacy taxonomy for data privacy policies. The *DataPrivacyPolicy* class incorporates other sub-classes which are called, *Purpose UsePrivacyPolicy* (data use), *VisibilityPrivacyPolicy* (data accessibility), *GranularityPrivacyPolicy* (data precision), and *RetentionPrivacyPolicy*, amongst others. Each of these entities describe different aspects of the privacy policy and offer unique modelling semantics for preserving data provider's private and sensitive data. We illustrate the broad description of the *DataPrivacyPolicy* entity for attribute data privacy preferences modelling in Figure **3**.

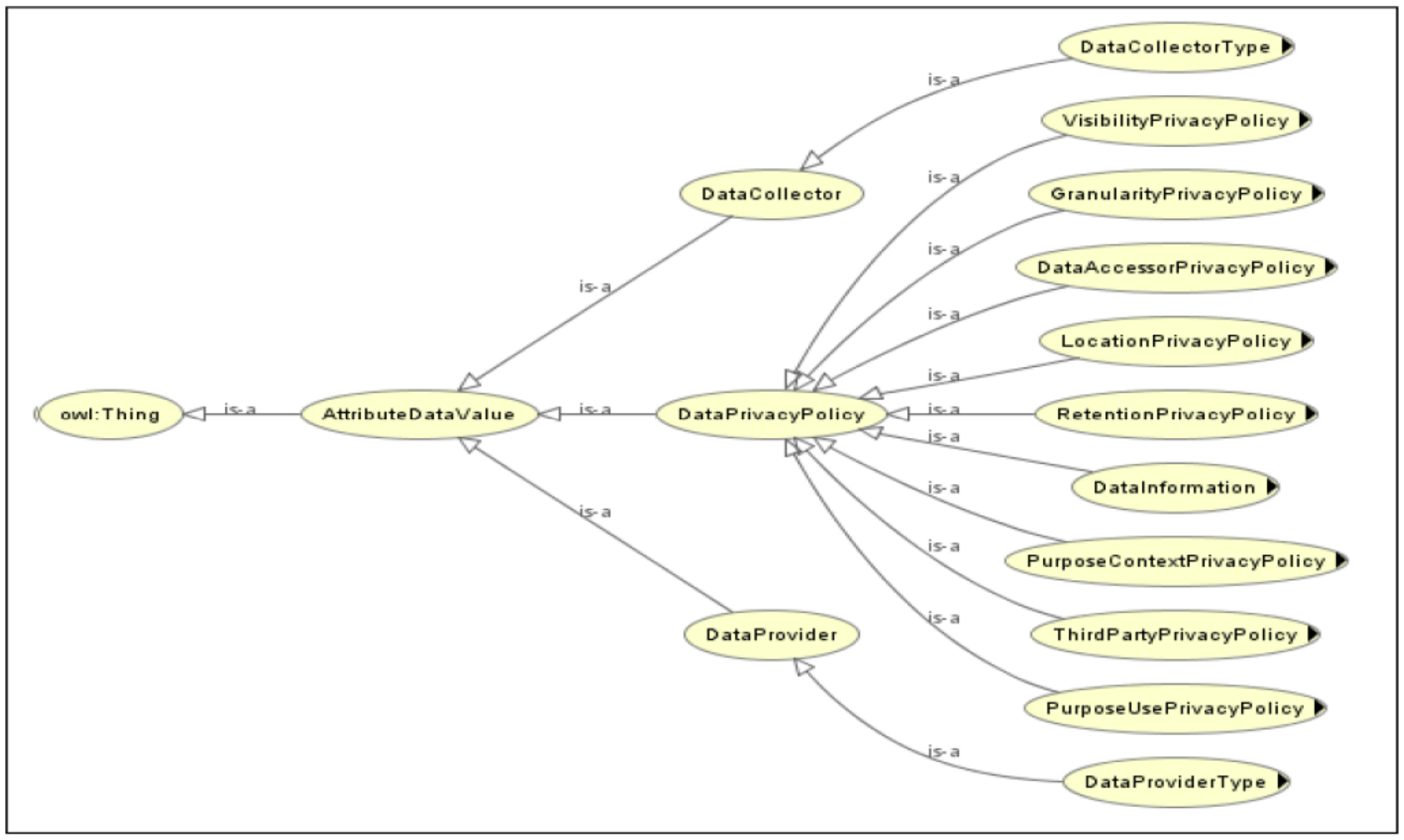

**Figure 2:** Formal Contextualized Privacy Ontology Model.

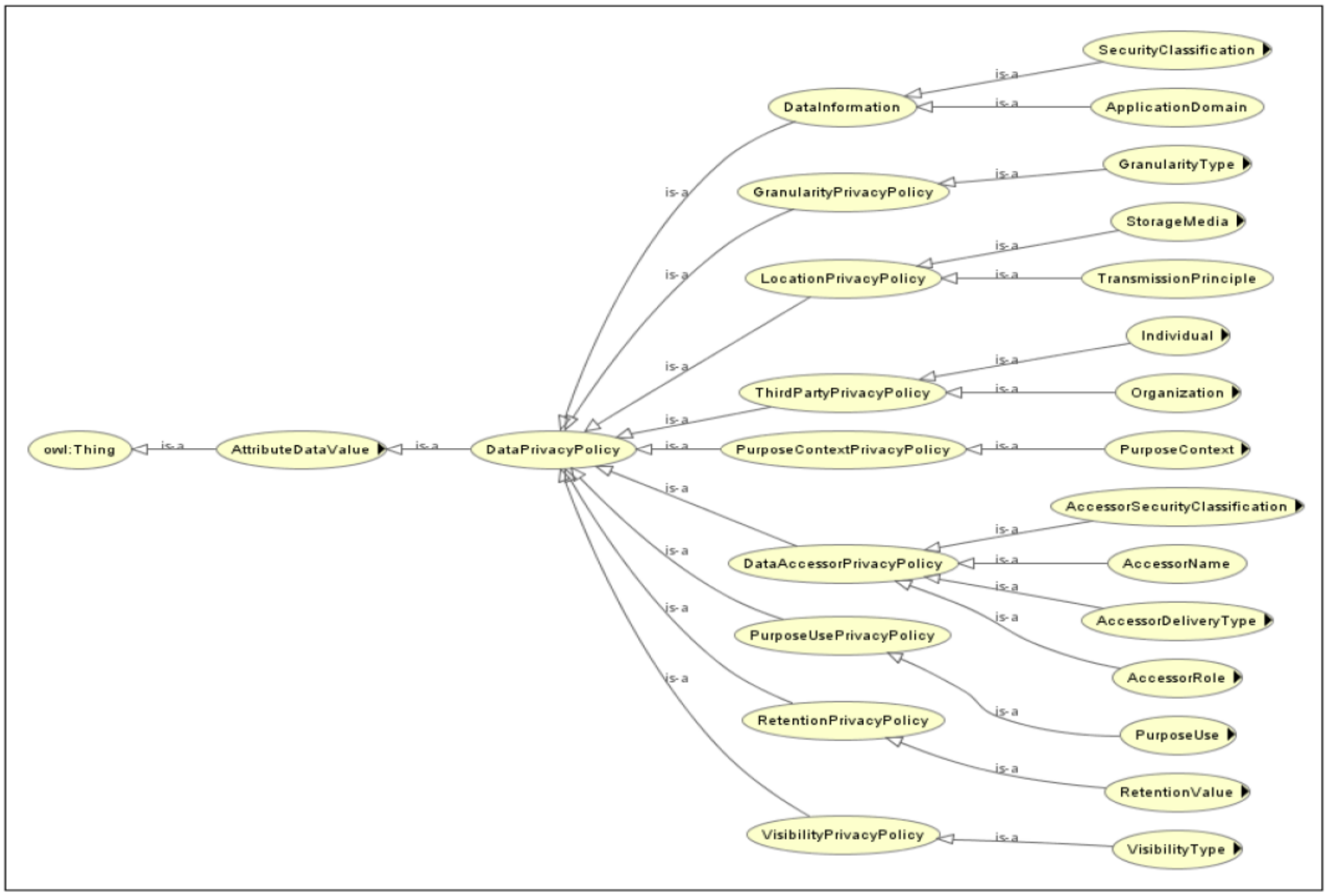

**Figure 3:** Attribute Data Privacy Preferences Ontology Model.

### *3.2.1. Privacy Ontology Update and Application Domain Adaptability*

The management of privacy policies and data provider privacy preferences involve regular change update by both data provider and collector. This is necessary because conditions regarding data provider's private data keep changing, and this will necessitate required updates or changes in existing privacy preferences. These procedures underlie and validate the dynamic formulation of the privacy ontology to handle diverse maintenance changes and its adaptability to different application domains. These update and maintenance procedures are expressed based on two instances.

First, it essential to update and subsequently remodel the privacy model (regarding new changes in the agreed privacy policy) for the overall private data management. This could arise in the context when third-party privacy policies are updated, retention duration is updated, or new data accessor role is defined, for some of the attribute data values. For example, in the event that new data access role needs to be defined for a data analyst. This implies that the ontology must evolve and change to align with the new attribute data value privacy preferences. Hence, as privacy preferences change, the privacy policy ontology model also changes.

Second, the privacy ontology evolves to support different application domain implementations. These expected maintenance changes may affect some structural aspects of the overall ontology design. For instance, the privacy ontology may be applied to a different application domain, such as, healthcare; and this will require some changes in the formulation of entity classes and their predicates. Thus, some predicates could be deprecated while new predicates are modeled to instantiate query processing and data retrieval within the context of current application domain.

In summary, we define the formulation of the privacy ontology as a generic (but dynamic) foundation for the privacy model in the overall privacy infrastructure methodology. Moreover, dynamic privacy ontology modelling allows verification by domain experts and to reflect entire privacy-aware data processing needs for an application domain.

### 3.3. Tight-Coupling of Data Elements

The data coupling component of the overall approach combines three data elements (namely, attribute meta-data, data provider privacy preferences data, and data accessor data) into a privacy tuple to be stored in the blockchains. In this approach, we bind all data provider's data elements for data privacy-preservation. This binding enables data retrieval or query processing on the attribute data to be dependent on the privacy preferences and data accessor profiles specified in the privacy policy and agreed upon by the data collector and/or accessor.

The *attribute data* ($d_i$) identifies each data provider private, sensitive attribute data collected from or provided by the data provider; as part of service account creation, personal biographical, demographic, and healthcare information, or data transactions performed by the data provider. Forms of attribute data collected are *birth date*, *phone number*, and *home address*, amongst others. In terms of the data provider privacy preferences, forms of data values collected are *purpose*, *granularity (data precision)*, *visibility (data accessibility)*, *retention duration*, *effective date*, *third-party data accessor*, and *purpose contextual norms* (of *when*, *who*, *why*, *where*, and *how*). It will be noted that these contextual privacy preferences are outlined based on the formal contextualized privacy ontology; which models and defines the overall contextualized data provider's privacy. A data provider *privacy preferences data* ($p_i$) is generated based on the data values extracted from the data provider's privacy preferences in the privacy policy.

Access to the underlying data values stored in the relational database is controlled and granted to data accessors with assigned access privileges. The assignment involves setting up different access levels, in accordance with the privacy policy statements. The determination of the data accessor profiling is based on factors and semantics, such as, *user role* (for *e.g.*, analyst, clinical physician), *data permission level* (for *e.g.*, view data, update data), and *data sensitivity level* (for *e.g.*, Level-2, Level-4), amongst others. A *data accessor profile data* ($u_i$) is generated to uniquely identify and grant access to which types of attribute data the data accessors can query or retrieve information.

The three data elements (*i.e.,* attribute data, data provider's privacy preference data, and data accessor data) are bound together to constitute the *privacy tuple*, $pt = < d_i, p_i, u_i >$. Hence, this privacy tuple is always preserved and respected whenever an attribute data value is to be retrieved from the relational database repository. Additionally, when the attribute data value is transferred to another data repository (for example, a third-party accessor repository), the commitment to the privacy policy tuple is respected and enforced. Consequently, access and data retrieval on the attribute data value is granted at the secondary data repository. The descriptive illustration of the tight-coupling of data elements is illustrated in Figure **4**.

### 3.4. Tight-Coupling of Relational Database and Blockchains

The data storage mechanism is based on the principle of coupling different database platforms: namely, relational database and blockchain. All data store components are hosted on a single computing machine, and this enables an efficient implementation of the data store coupling. We explain data store tight-coupling to indicate that privacy-aware data and query references are stored on the blockchain platform; and these are directly related to the content data stored on the relational database. These privacy preferences determine, authorize, and authenticate query processing and data retrieval of privacy-aware data records (from the relational database) that need to be presented to the intended data accessor. Hence, data processing on the relational database is always dependent (or tightly-knit) on the privacy preferences stored on the blockchains. Subsequently, this approach underlies no direct, independent query access or data retrieval from the relational database.

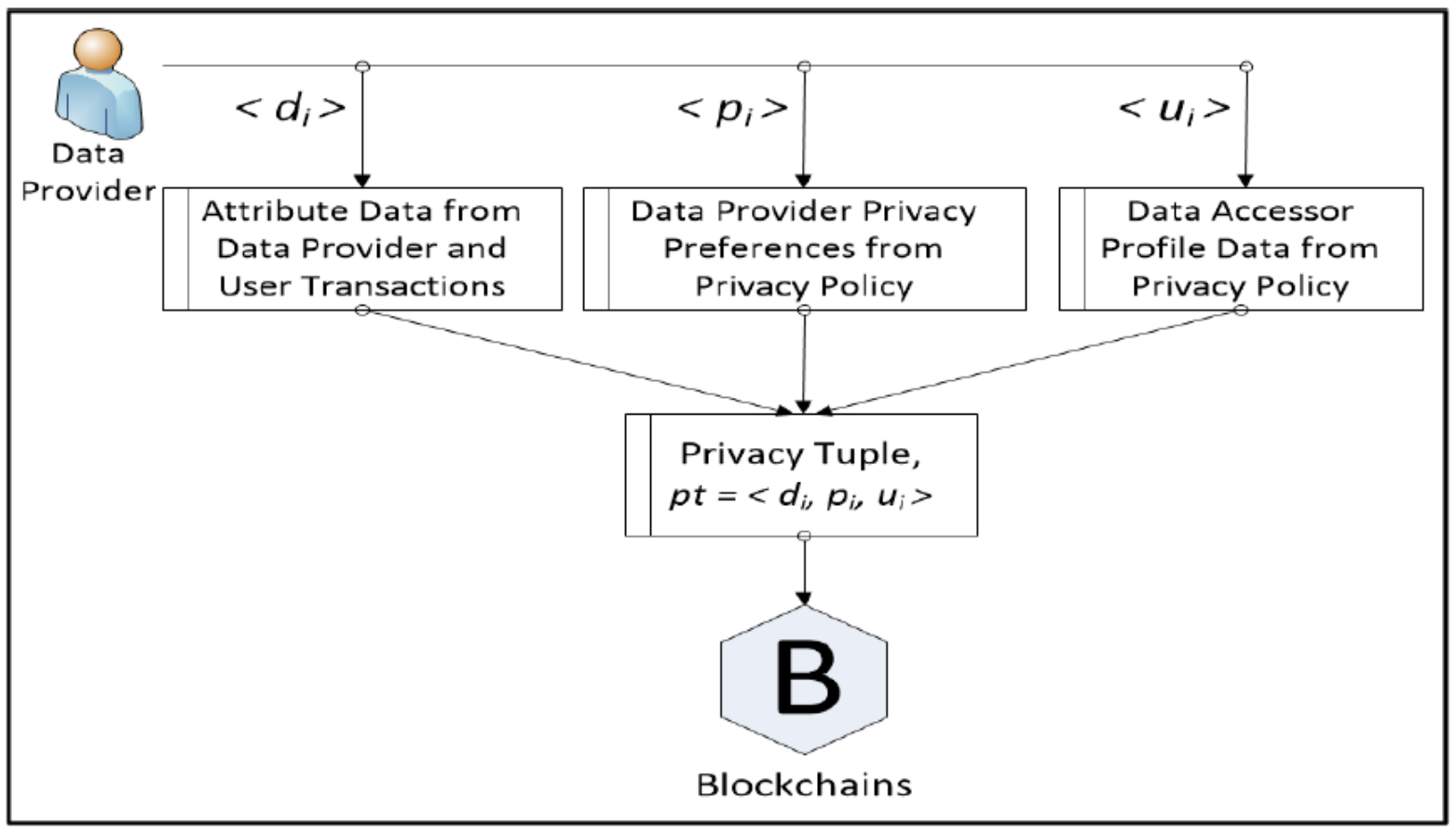

**Figure 4:** Tight-coupling of Data Elements into Privacy Tuple.

The reason for this approach is to maximize the merits of managing and processing transaction data from each of the data platforms. The relational database offers a scalable, high throughput, and efficient querying engine to meet expected high data processing needs. In contrast, the blockchain platform offers decentralized data control that is tamper-resistant and immutable for the protection of privacy-related data values. Additionally, decentralized control of the blockchains offers efficient change request and approval (from service providers and data providers). The blockchains offers a platform that can be used to facilitate privacy audit procedures for data provider privacy preferences. One key advantage for the incorporation of the blockchains in the privacy model is the ability to provide a platform for all data accessors (data and service providers) to agree on the state and value of data stored. Additionally, the blockchain platform provides an infrastructure platform to effect efficient changes to privacy-related data values. The tight-coupling approach also enables either input of data provider and service provider in the agreement of the parameters for privacy policies; thereby ensuring access consistency for all data accessors in a reliable data provider privacy and identity management.

The high-level methodology is, as follows: the entire set of data values (on data provider personalized information and transaction data) are collected and stored in the relational database. Moreover, an instance of the integrated privacy tuple data is stored in the relational database. On the blockchains, another instance of the tightly-coupled privacy tuple and privacy preferences (which is subsequently hashed) are stored. Figure **5** illustrates the proposed architecture of the relational database and blockchain. With this data platform integration, every data retrieval request from the relation database is authenticated and authorized from the blockchain platform (through its related transaction ledger). This establishes the tight-coupling approach of the integrated data platform, and no other data retrieval (on the relational database) is performed independent of the blockchain. The data and communication connection between the relational database and blockchains is via an encrypted data communication protocol. Furthermore, any form of data transfer (query requests and results) between the data repositories is completed in a secured, tamper-resistant access-controlled protocol.

### 3.5. Query Processing on Proposed Methodology

The proposed methodology takes query requests to the relational database by first mediating them to the blockchains platform before delivering query results. We describe this procedure as privacy-aware; as the query data elements and the expected query data results or tuples are accessed and verified through the privacy preferences of the data provider. The query request is first transmitted to the privacy-aware query analyser. The privacy-aware query analyser parses the query, identifies and extracts attribute data (regarding the requested query) that needs to be accessed for data retrieval. The privacy-aware query attribute data are then transmitted to the blockchains.

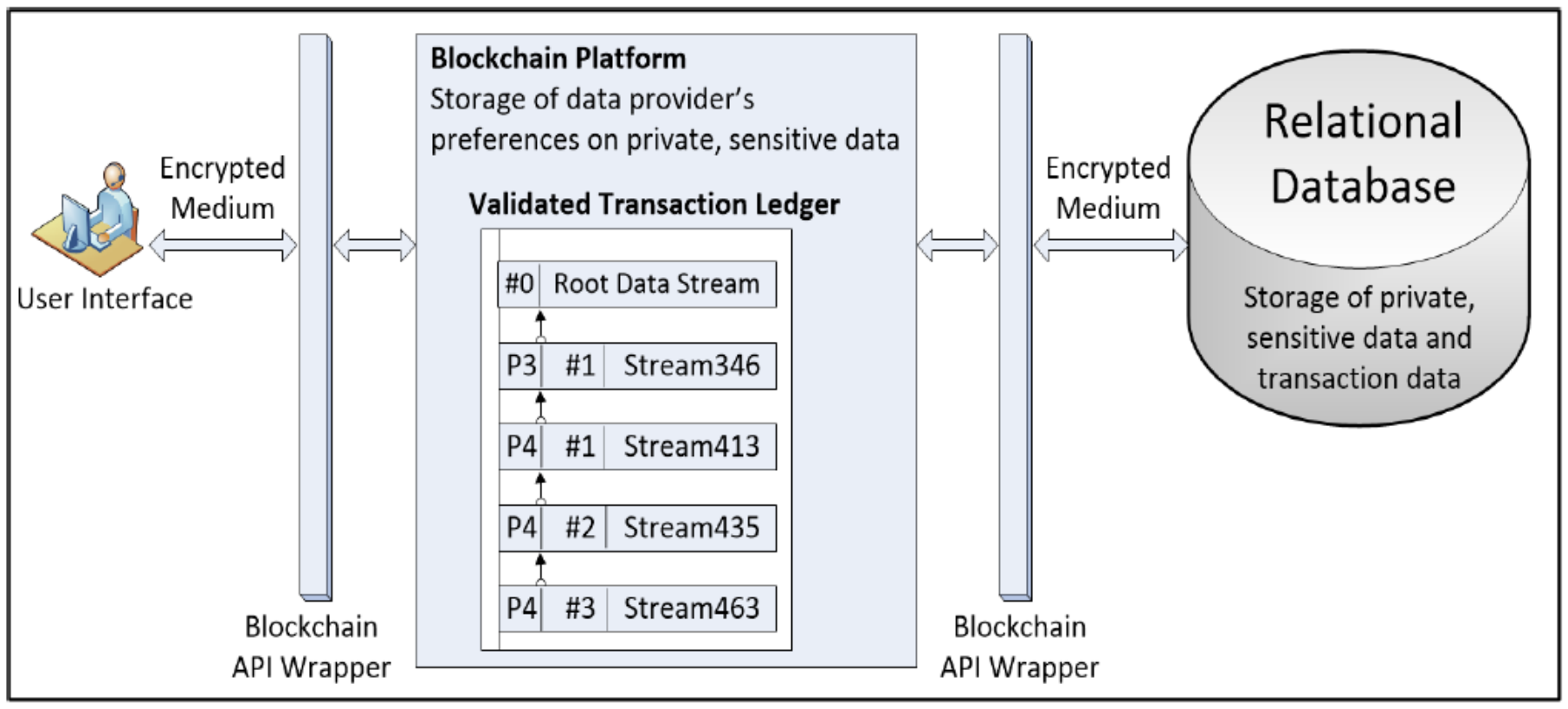

**Figure 5:** Architecture of Tight-Coupling of Relational Database and Blockchains.

At the blockchains, the data provider's transaction ledger data stream needed for data retrieval is identified. The blockchain platform traverses through the transaction data stream to access the most *recent* privacy tuple transaction data item block. Transaction data item block corresponding to privacy tuple is subsequently accessed and retrieved. This data item contains *recent* data provider privacy policy preference's tuple (on attribute meta-data, contextualized privacy preferences data value, data accessor value, and other privacy preferences).The blockchain API delivers retrieved data items (in its hashed form) to the privacy-aware query processor.

At the query processor, the privacy-aware query is rewritten and transformed. Data items consisting of privacy policy preferences are verified against the retrieved hashed data values from the blockchain. In cases where there is data inconsistency between both data item values, the query is denied and aborted. By data inconsistency, we identify that the hashed data values stored in the blockchain is different from data values retrieved from the relational database. Upon valid consistency check on data item values, the query processor grants access for query processing and data retrieval. The privacy preferences on each data attribute are used to rewrite or transform the privacy-aware query to conform to the current context based on data provider preferences in the privacy policy.

The privacy-aware query processor uses the privacy-aware rewritten query statement to generate expected query result or data instance tuples. Consequently, data instance tuples regarding the attribute data are retrieved and presented to the data accessor through user interfaces. In cases where a new preference is consented and approved by the data provider (that is, user) through the system user interfaces, a new data is written and saved on the blockchain. This new preference record becomes the latest retrievable privacy preference for all future record query processing. Moreover, in terms of consent withdrawal, no query is processed and the subsequent requested query to the data accessor is denied or aborted. A consent data item saved in the blockchain prevents any privacy preference record from being retrieved.

In summary, the blockchain platform acts as a privacy-aware data access control and query analysing medium to the underlying relational database. Hence, whenever a data accessor wants to query or retrieve data values in the underlying relational database, access is routed to the blockchains before privacy-aware data values stored in the relational database are retrieved and presented to the data accessor.

Figure **6** illustrates the implementation of the query processing architecture for accessing data provider private, sensitive data from the relational database repository. The architecture presents an integrated query platform of user interface, relational database, and blockchain. The user interface provides the

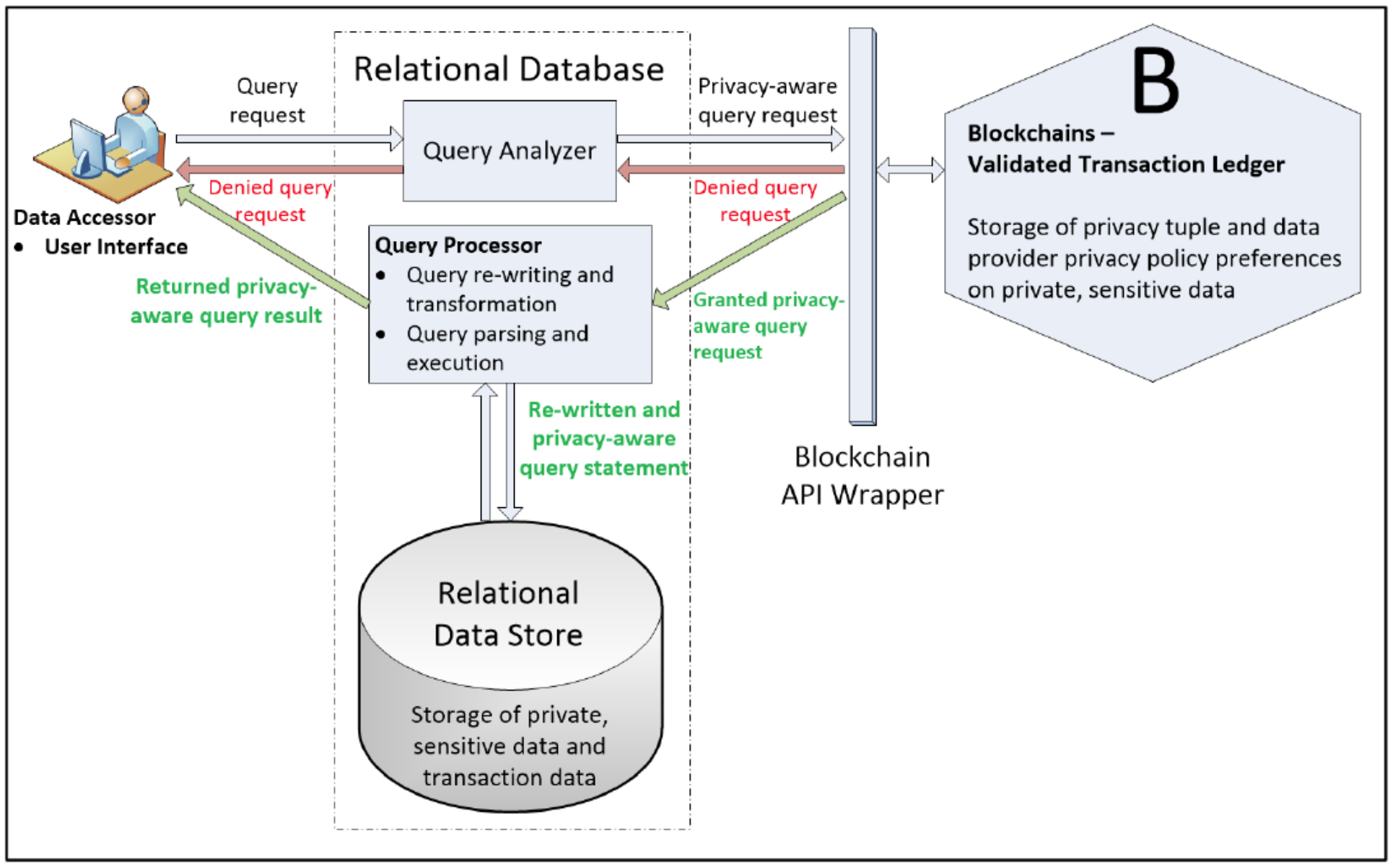

**Figure 6:** Query Processing on Relational Database and Blockchains.

medium for data accessors to post queries and display returned query results. The relational database is made up of the data store, privacy-aware query analyzer, and privacy-aware query processor. The data storage stores all private, sensitive and transactional data. The privacy-aware query analyzer is designed in the form of a database stored procedure, that accepts data parameters to allow or reject queries. The privacy-aware query processor executes the queries. The blockchain is composed of validated transaction ledgers for data provider privacy tuples access and retrieval information.

## 4. RESEARCH EXPERIMENTAL IMPLEMENTATION

We implement the research methodology by modelling, designing and developing different system components. These are privacy ontology, user interfaces, relational database, and blockchains platform.

### 4.1. Privacy Ontology Formulation and System Modelling

We model and design several entities, sub-classes and their predicate classes. Each entity contributes to the overall modelling of the privacy policy regarding a particular data attribute or category of attribute data. This privacy ontology outlines and details every aspect of the privacy policy for each attribute data, and the procedures for data storage and management. We implement privacy ontology modelling using Protege Semantic Web Ontology [30] application tool.

### 4.2. User Interface Modelling and Development for Privacy Tuple

The overall user interface system architecture involves modelling and design of user interfaces to collect and display data to the actors of the system. As part of data collection, the data provider offers data privacy policy preferences which are validated by the data collector (or service provider). Additionally, the data provider produces private, sensitive transactional data. Other related data are collected by data collectors alongside the privacy preferences data. These data privacy policy preferences and private, sensitive data are transferred from user interfaces and stored in the relational database for data processing and report generation. Furthermore, a hashed form of the privacy preferences and privacy tuple is saved on the blockchain platform.

We modelled and designed the user interfaces using Unified Modelling Language (UML) use case diagrams. Use case diagrams show interactions between the system and its environment. We developed and programmed the user interfaces using PHP (PHP: Hypertext Preprocessor) web development tool.

### 4.3. Relational Database Design

We adopt MySQL database management system as the relational database. The database platform provides functionality for data storage, management, query processing, and native data objects, such as, tables, indexes, and stored procedures. Additionally, the platform provides data processing on data provider biographical and demographic information and other entities.

Three different databases are created in the relational database integration. These are *user_admin*, *tunote_ppdb*, and *tunote_data_provider*. The *user_admin* database serves to process and store data on system user login details and access privileges. This database contains 2 relational tables. The second database, *tunote_ppdb*, processes and stores data on data provider privacy policy preferences, such as, *attribute category data*, *data accessor profiles*, and *granularity privacy*, amongst others. The database contains 16 relational tables. Some of the table information are; *attribute_category_data* (8 records), *attribute_data* (36 records), *data_accessor_profile* (5 records), *etc.* The third database, *tunote_ngs_ genomics*, processes and stores content data relating to data provider information, physician information, healthcare requisition details, and medical consent details. The database contains 9 relational tables, such as, *data_provider*, *physician*, *privacy_consent*, amongst others. Some of the table information are; *privacy_consent* (15 records), *patient* (3000 records), *physician* (20 records), *requisition* (50 records), etc. The query processing functionality (which involves the query analyzer, processor, and execution) of the system is implemented using a stored procedure database object. The stored procedure acts as a privacy-aware API over the database to process (that is, query analyses, parsing, and transformation) and execute queries. The stored procedure accepts parameters in the form of data provider privacy preferences object identifier values and present privacy-aware query results to the data accessor. The output from the stored procedure is a set of privacy-aware data tuples based on the input parameters. In this way, the privacy-aware stored procedure offers functionalities of data validation and access-control mechanisms.

### 4.4. Blockchains Design and Development

We adopt a blockchain platform to offer authentication and authorize query processing on data provider private and transaction data, respectively. We employ MultiChain blockchain [39]; which is a robust platform for private blockchain development. MultiChain blockchain is an open-source, "off-the-shelf" platform for creation and deployment of private blockchains, either within or between organizations. MultiChain is designed to offer functionalities which: (a) ensures that the blockchain's activity is only visible to chosen participants, (b) introduces controls over which transactions are permitted, and (c) enables mining to take place securely without "proof-of-work" (PoW) and its associated costs.

#### *4.4.1. Design of Transaction Ledgers and Data Streams for Privacy Tuples*

We created a *chain* transaction ledger and a number of data streams. Data streams are append-only, on-chain lists of data - where the *key-value* pair retrieval capability makes data storage and query functionality extremely easy. Each data stream is an ordered list of items, with the following characteristics: (a) one or more *publishers* (string data type) - who digitally signed the data item; and (b) one or more *key-value* pairs (string data type) - which is set between 0-256 ASCII characters excluding whitespace and single/double quotes - to allow efficient retrieval.

MultiChain data streams make it possible for a blockchain to implement some basic data management features of a general-purpose database. Storing data on the blockchain involves the process of *publishing* data. The data published in every stream is stored by all nodes in the network. Each data stream on a MultiChain blockchain consists of a list of data items. When data needs to be queried or "streamed", it can be retrieved by searches using the *key-value* pairs. Publishing a stream item to a data stream constitutes a transaction. When a node subscribes to a data stream, it indexes the stream items in different ways to enable fast retrieval, and the index entry points to the transaction ID. The core value of data streams is in indexing and data retrieval. The node address *subscribes* (with a granted permission to retrieve transaction data) to the created data stream. After data stream subscription, the node can retrieve data items in a specified order and retrieve data items with a particular *key*. Our configuration of the blockchain API wrapper use *start* and *count* parameters; that allows subsections of long lists to be efficiently retrieved (like a LIMIT clause in SQL).

For each data stream, we create validated transaction data block (associated to a privacy tuple). The initial configuration of the blockchains consists of appending a *root (genesis)* data stream that does not contain transaction data block. Consequently, the first data stream is appended to the existing *root* data stream, and all other subsequent data streams are also appended to the preceding data stream. A validated transaction data block (associated to a privacy tuple) is appended to first data stream or any other related data stream (in the list of data streams created on the chain). Thus, privacy tuples stored as transaction blocks become valid data items and are always accessed during query processing.

The implementation approach for designing and creating data streams is as follows. We adopt a data representation format of: *Stream-X-Y-Z*. The *X* component represents a data value for supposed data provider, *Y* represents a data value for supposed attribute data, and *Z* represents a data value for the data accessor. This approach offers a unique data stream creation and identification on MultiChain blockchain. Furthermore, the methodology offers an efficient privacy tuple data storage for each data provider, the associated attribute data, and the data accessor. In terms of data retrieval and query processing on MultiChain blockchain platform, we reference the parameters using object identification data values to uniquely identify data streams created on the blockchain. We use MultiChain API syntax: *create ("stream", stream_name, True)*; to create a data stream from the API. The parameter *"create"* indicates the method invoked to execute processing on the blockchain platform. The *"stream"* parameter indicates the type of operation, whether data stream or wallet asset. The *"stream_name"* indicates the specific name assigned to the data stream.

Figure **7** illustrates a set of subscribed data streams, containing the *"root"* stream (which is automatically created) and user-created streams of *"stream-4-5-3"* and *"stream-4-1-3"*. It is noticed from the diagram that all three data streams are created by the same node address. Data streams *"stream-4-5-3"* and *"stream-4-1-3"* are designated to the same data provider with an object identifier value, *4*. On the other hand, each of the streams are associated with different attribute data with an object identifier value of *5* and *1*, respectively. Finally, each data stream is related to the

Subscribed streams

| | |
|---|---|
| **Name** | root |
| **Created by** | 1ZPmVpyGqc3J9g8gsiVMEMnzUy9hqTFhQJRNTf |
| **Items** | 0 |
| **Publishers** | 0 |
| **Name** | stream–4–5–3 |
| **Created by** | 1ZPmVpyGqc3J9g8gsiVMEMnzUy9hqTFhQJRNTf |
| **Items** | 3 |
| **Publishers** | 1 |
| **Name** | stream–4–1–3 |
| **Created by** | 1ZPmVpyGqc3J9g8gsiVMEMnzUy9hqTFhQJRNTf |
| **Items** | 4 |
| **Publishers** | 1 |

**Figure 7:** MultiChain Blockchain (Subscribed) Data Streams.

same data accessor with an object identifier value of *3*. Moreover, *"stream-4-5-3"* stores three data items and has one node address data publisher. Likewise, *"stream-4-1-3"* stores four data items and has one node address data publisher.

#### *4.4.2. Publish and Confirmation of Privacy Tuple Data Items on Data Streams*

*Publishing* data is the methodology for saving and storing data value items on the MultiChain blockchain. The process of publishing data value items involves identifying a data stream (which is already created and related to a unique data provider, attribute data, and data accessor), specifying a key value, and specifying the data value item.

As part of implementation, we adopt a hashing algorithm to transform privacy tuple data items before finally saving them on the blockchains. We implement native hash algorithm offered by PHP Interpreter for web development. We use *bcrypt* algorithm (in the form of *Argon2i*) to transform the privacy tuple and privacy preferences data. The resultant hashed value contains the hash algorithm, hash cost and salt. Therefore, all information that is needed to verify the hash is included in the resultant hash value. For practical implementation on the MultiChain blockchain, we use MultiChain API syntax: *publish (stream_name, stream_key, stream_data)*; to publish or store a data item in the data stream. The parameter *"publish"* indicates the method invoked to execute processing on the blockchain platform. The *"stream_name"* indicates

Stream: stream–4–3–5 – 4 of 4 items

| | |
|---|---|
| **Publishers** | 1ZPmVpyGqc3J9g8gsiVMEMnzUy9hqTFhQJRNTf |
| **Key(s)** | key18 |
| **Text data** | $2y$10$WKq1AB.HZvzapZACWHZLX.ZUHODgpZa65yMugkCpWxUdBzqv5/BKi |
| **Added** | 2021-04-02 20:21:54 GMT (confirmed) |
| **Publishers** | 1ZPmVpyGqc3J9g8gsiVMEMnzUy9hqTFhQJRNTf |
| **Key(s)** | key18 |
| **Text data** | $2y$10$Z5MdS8N22hgK3D/QBOnFU.Z0TUCerxzZAZjPe6.2z1rz5Qv/ZYQTe |
| **Added** | 2021-04-02 20:13:39 GMT (confirmed) |
| **Publishers** | 1ZPmVpyGqc3J9g8gsiVMEMnzUy9hqTFhQJRNTf |
| **Key(s)** | key18 |
| **Text data** | $2y$10$kt/OVpH20JKb2/lYCjxW0eCF5nZ5TTBJYMNl9LBfD/F8g2t5HIInS |
| **Added** | 2021-03-21 06:59:34 GMT (confirmed) |
| **Publishers** | 1ZPmVpyGqc3J9g8gsiVMEMnzUy9hqTFhQJRNTf |
| **Key(s)** | key17 |
| **Text data** | $2y$10$5OxvVxggEPxb6lsIpxupnOXz58zbC.oqmZ4pDYwMbe1b5JO0IzdTG |
| **Added** | 2021-03-21 06:52:32 GMT (confirmed) |

**Figure 8:** MultiChain Blockchain Data Stream Items -Data Provider Privacy Tuple.

the specific name data stream into which data is stored, and *"stream_key"* parameter indicates the data value of key that is used to sign the storage of data item on the data stream. The *"stream_data"* indicates the actual data value item saved on the data stream.

Figure **8** illustrates the set of data value items for *"stream-4-3-5"*. It is observed (from the timestamp) in the illustration that ordering of the data value items is in an ascending order from the bottom. Hence, the latest data value item is displayed at the top. We notice that all the data storage streams are published by the same node address. Moreover, the first data value item is signed with a key value of *"key17"* whereas all subsequent data value items are signed with *"key18"*. It is also noted that all the data value items are "confirmed" by the subscribing nodes to the data stream. This indicates data value item has been accepted as a valid data from each subscribing node.

### 4.5. Data Communication and Encryption

Data communication between system components are facilitated using encrypted channels. We adopt secured application and transport protocols to provide data security over the communication channels between system components. For robust and efficient data encryption, we adopt TLS (Transport Layer Security) protocol - with TLS_AES_256_GCM_SHA384 cipher suite - to ensure secure data communications for all forms of data packets transmission between system components [31, 40]. Moreover, data communication between the blockchains, relational database and user interfaces is facilitated using an API wrapper. The blockchain API wrapper is implemented in PHP programming platform.

We adopt a PHP string encryption with the *libsodium* library. This library is easy-to-use software collection for encryption, decryption, and signatures. The encryption schemes implemented in *libsodium* are fast and require a unique (nonce, key) tuple for every plaintext. This encryption methodology is considered efficient against side-channel or server-side attacks [41, 42]. The implementation with Libsodium encryption adopts different algorithms for different purposes of: encryption/data integrity, AES key agreement, and source/entity authentication. For encryption/data integrity, the encryption library employs XSalsa20 to encrypt stream cipher and AES-GCM-256 is used for key agreement and for message-authentication code (MAC) in one operation. The adopted encryption algorithm provide *confidentiality* of the data message's contents, *verifies* the data message's origin, and provides proof that a data message's contents have not changed since it was sent (*i.e.*, *integrity*). Hence, the encrypted communication medium sufficiently reduces (or best case prevents) any form of attack, data loss, or data leakage as data moves between the system components [31].

## 5. EVALUATION AND RESULTS ANALYSIS

We evaluate and analyse the output from the proposed privacy infrastructure based on the propositions discussed in Section 3 and implementation procedures discussed in Section 4.

### 5.1. Security / Adversarial Model

We present the overview of the security model which addresses the design of secure schemes or protocols. Figure **9** illustrates the security model design and adversarial model components: assumptions, adversaries, goals, and assumed capabilities. Additionally, Figure **10** illustrates security/adversarial model components for the overall security system design.

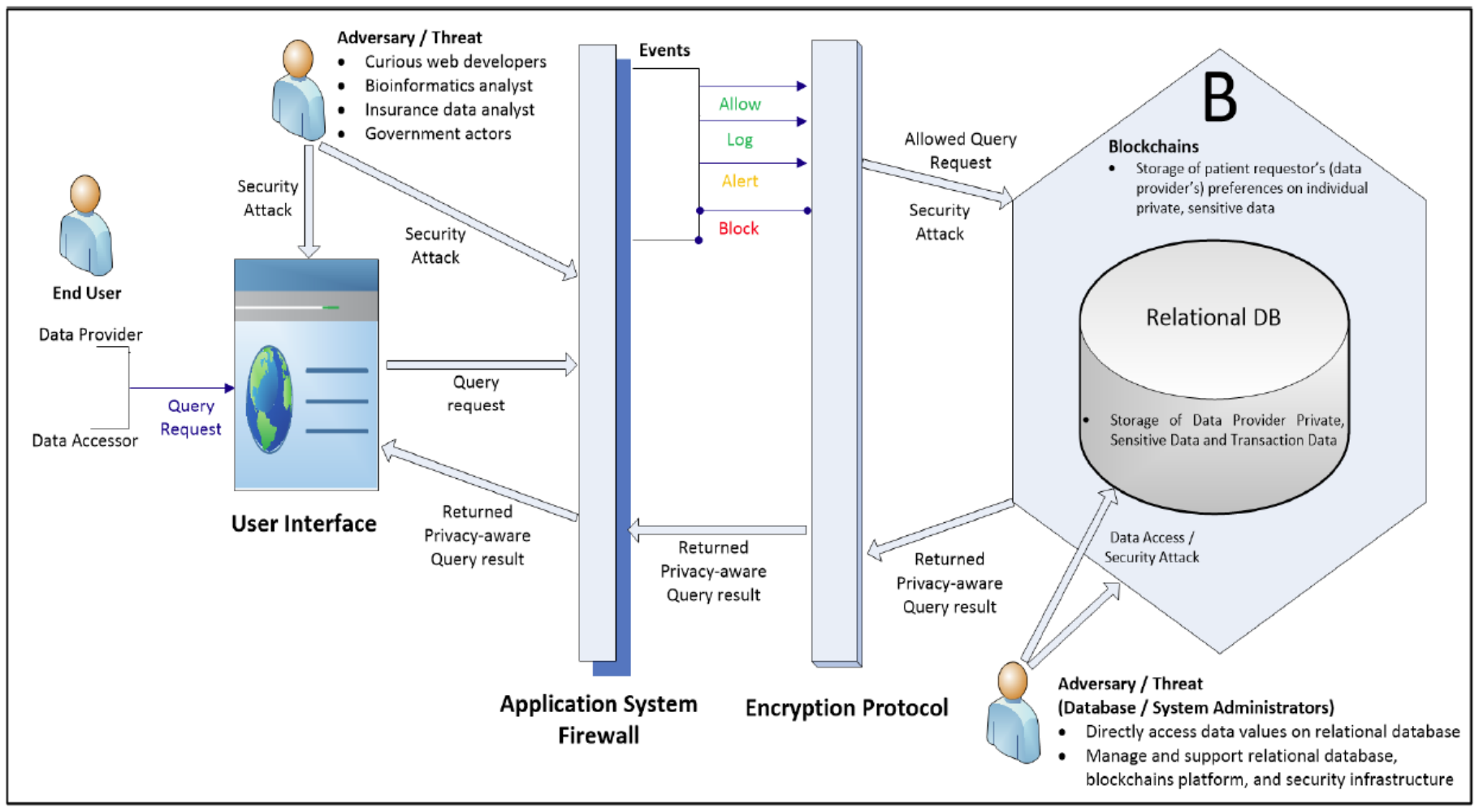

**Figure 9:** Adversarial Model System Design.

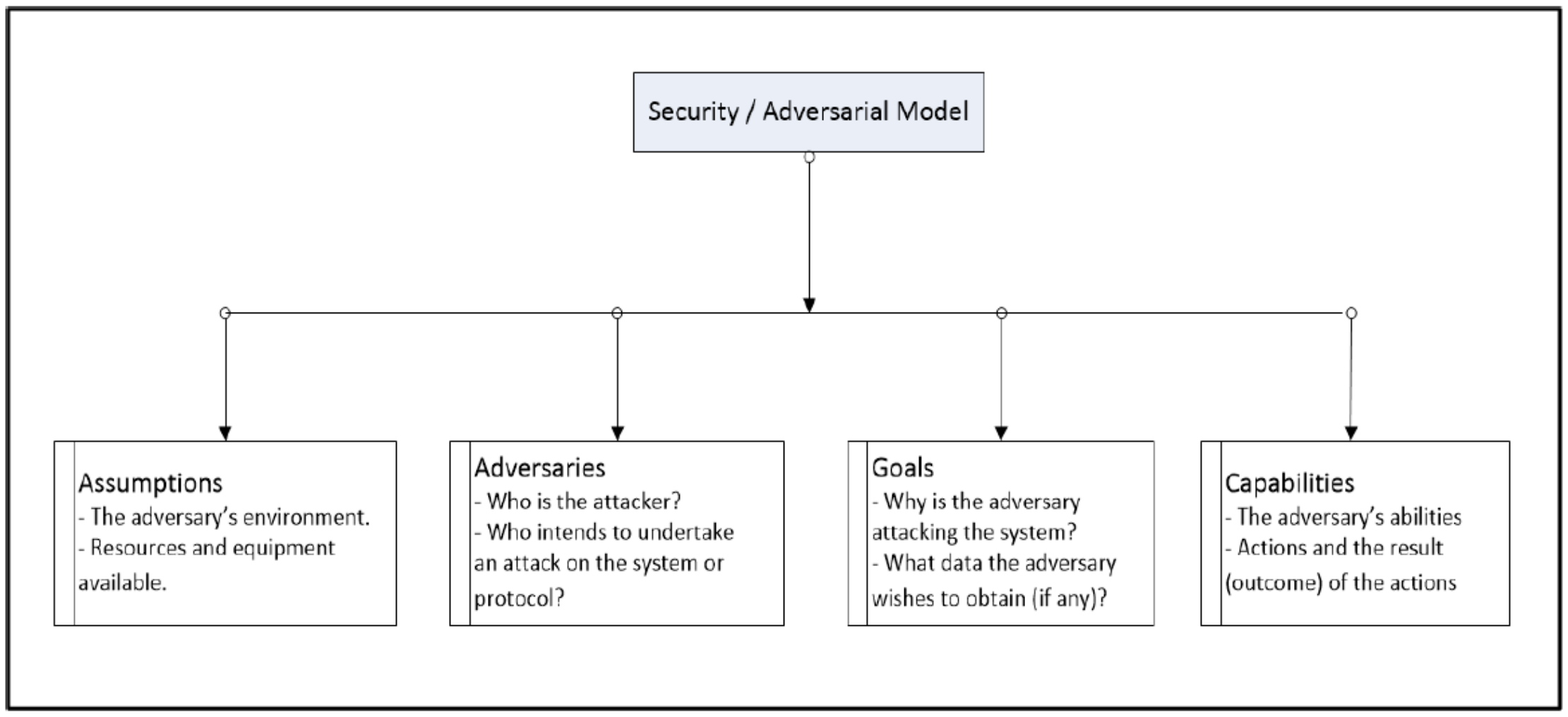

**Figure 10:** Adversarial Model Components.

**Assumptions:** The assumptions framed for the adversarial model is based on the environment and resources available. In terms of the environment, the primary platforms and devices are: the user interfaces, encrypted local network, relational database, and blockchains platform. Some resources available to the adversaries are: access privileges to the user interface, relational database, blockchains platform, knowledge of tight-coupling of relational database and blockchain.

**Adversaries:** An adversary to the system is described as an individual or a system (whether authorised or unauthorised) that tries to access data from either the relational database or blockchains platform, or both. Some end users who serve as system threats are the database administrator and system administrator. Adversaries may be categorised as either active or passive threats.

**Adversary Goals:** The primary goal of the adversary is to have the ability to access data values in the relational database without querying the blockchain platform; which serves as an access control authentication medium. Additionally, the adversary may try to make changes to the privacy tuple generated for a data provider (by either changing the data values for data provider, privacy preferences, or data accessor). More importantly, the adversary may try to modify existing privacy policy preferences entries, for example, granularity privacy, visibility privacy, retention duration, *etc.* The adversary may try to access the transaction ledgers and data streams on the blockchain platform. The adversary may try to query hashed privacy preferences data values.

**Assumed Capabilities:** In terms of the security infrastructure, the systems administrator or network engineer (or other active adversary) may have *listen* capability that allows her to track or eavesdrop into data packets flowing through the application network medium. On the other hand, the database administrator may have the access to read (or query) or modify data values from the relational database or blockchain platform.

### 5.2. Privacy Infrastructure Security Assessment

We discuss the privacy infrastructure systems security overview in terms of the defences it offers for data provider's private data and overall data processing for data accessors. We discuss these assessment based on the following: *confidentiality*, *integrity*, and *availability*.

We address *confidentiality* in which the privacy infrastructure protects private data from unauthorized access using defined data provider privacy policy preferences saved on the blockchain platform. Moreover, the composition of the privacy tuple authorizes data access to each defined accessor based on privacy preferences. In terms of *integrity*, our proposed methodology offers a framework where private, sensitive data and the preferences defined on them are protected from deletion or modification from unauthorized individuals. Our assessment ensures that changes to privacy preferences stored in the blockchain platform is completed in "consensus" or agreement by both data provider and data collector.

We address security assurance of *availability* based on the tight-coupling of data storage and processing platforms. The architecture and configuration of the blockchain nodes offer functional transaction ledgers and data streams that ensure controlled data throughput to the API and relational database. The relational database is configured with stable query analyzers and processors to parse and execute query statements; and to deliver data values to data accessors.

### 5.3. Query Processing and Response Time Analysis

We analyze query processing rate on the privacy model system implementation to evaluate the effectiveness of running data queries. The data storage and query processing is done on a single computing machine with eight logical processor(s), 3.6 GHz processing speed, and 16 GB of RAM, and 900 GB of disk storage capacity. We identify three different types of queries (on data provider demographic data, healthcare data, and consent witness data), and discuss the results output from running these queries. Each query presents a unique set of attribute data elements, as well as different set of data provider privacy preferences. We run random independent queries for both privacy-aware and native queries for each query type. For each query type, we run 20 random repeated queries for each native and privacy-aware query, respectively. We then compute averages for each native or privacy-aware query form. The average query response time for each native and privacy-aware query on formulated query type is computed and the privacy overhead cost is calculated. As a result, we effectively compare the query results and establish the effectiveness and efficiency of the privacy-aware query medium.

#### *5.3.1. Query 1 - Data Provider Demographic Data*

Suppose we want to retrieve data values on data provider demographic information. The attribute data elements from which the query transaction is performed, include: *Street Name*, *City*, *Province*, *Postal Code*, *Original Province*, and *Phone Number*, *etc.* The result of query processing rate for data provider demographic information is illustrated in Figure **11**. The privacy-aware queries and native queries are designated with blue and orange colouring, respectively.

**Results Analysis and Discussion:** In Figure **11**, we observe an increase in query processing time for the privacy-aware queries in comparison to native queries. This is expected because of the added activity of privacy-aware query parsing, processing, and execution which leads to an increase in query data retrieval load or query cost. From the query results, we realize the highest and lowest query processing time for native queries are 19.01 ms (milli seconds) and 11.52 ms (milli seconds), respectively. Moreover, an average query processing time of 13.37 ms is attained. In terms of privacy-aware query processing, we attain highest and lowest query processing time of 24.86 ms and 19.48 ms, respectively. We also record an average privacy-aware query processing time of 22.99 ms.

We observe privacy overhead cost of 41.85% for privacy-aware query processing. This is indicative of the varied number of demographic attribute data and the instance data stored on these attribute data. More importantly, we analyze the privacy preference parameters of data information privacy with instance data representation of: *'Level-4: Restricted'*, granularity privacy with instance data representation of: *'Specific: Specific data item is accessed. A query for data item returns actual data value'*, visibility privacy with instance data representation of: *'Third-Party Allied Health Access: Data accessible to parties not covered by explicit agreement with House'*, and purpose privacy with instance data representation of: *'Reuse-Same: Data used for same purpose and multiple times'*. This analysis by the privacy-aware query analyzer increases the query overhead cost for the overall query processing. Based on the query response times attained from both privacy-aware and native query processing, we determine that query processing on the proposed privacy model architecture is efficient, and there is safe, cost-effective data retrieval on the overall architecture.

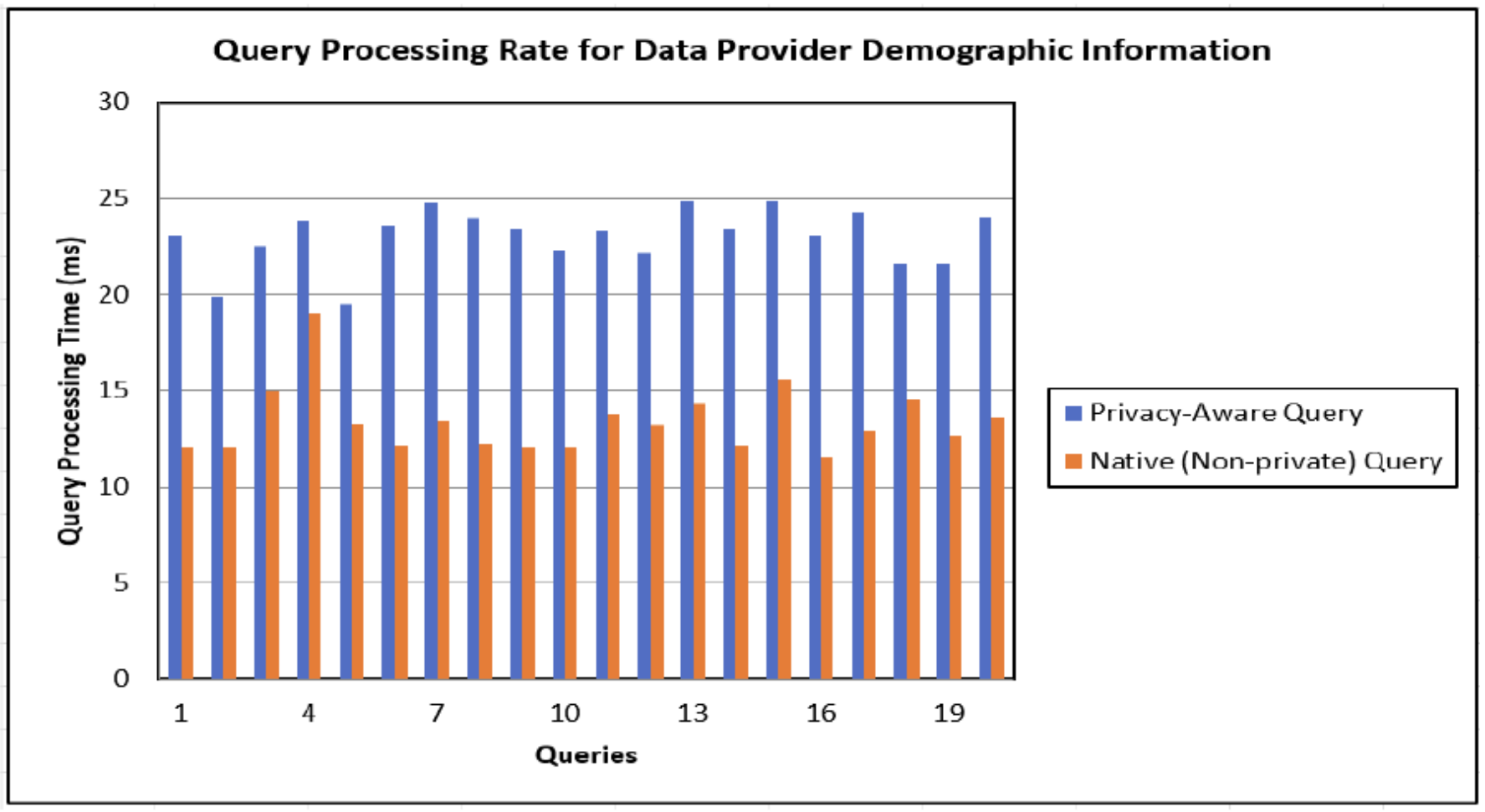

**Figure 11:** Query Processing Rate for Data Provider Demographic Data.

### *5.3.2. Query 2 - Data Provider Healthcare Data*

Suppose we want to retrieve data values on data provider healthcare information. The attribute data elements from which the query transaction is performed, include: *Personal Health Number*, *Medical Record Number*, *Chart Number*, *Personal Care Physician Name*, and *Dentist Physician Name*, *etc.* The result of query processing rate for data provider healthcare information is displayed in Figure **12**. The privacy-aware queries and native queries are designated with blue and orange colouring, respectively.

**Results Analysis and Discussion:** Similar to *Query 1*, *Query 2* also shows an increase in the query processing time for the privacy-aware queries in comparison to the native queries. In terms of native (non-private) query processing, we attain highest and lowest query processing time of 18.38 ms and 9.74 ms, respectively. We also record an average query processing time of 13.38 ms. Regarding privacy-aware query processing, we note that the highest and lowest query processing time are 23.97 ms and 19.03 ms, respectively. Moreover, an average privacy-aware query processing time of 21.67 ms is attained.

In terms of privacy overhead query cost, we observe 38.26% for privacy-aware query processing in relation to native (non-private) query processing. It will be noted that data provider healthcare information processed here involve unique set of privacy-aware attribute data, which are not many in comparison to demographic data in *Query 1*. Moreover, the privacy overhead query cost is attributed to data provider privacy preference parameters of data information privacy with instance data representation of: *'Level-3: Confidential'*, granularity privacy with instance data representation of: *'Partial: Partial or altered and non-destructive data values returned to accessor'*, visibility privacy with instance representation value of: *'Third-Party Allied Health Access: Data accessible to parties not covered by explicit agreement with House'*, and purpose privacy with instance data representation of: *'Reuse-Selected: Data used for primary purpose for which data'*. Finally, we note that query response times attained from both privacy-aware and native query processing determine that proposed privacy model

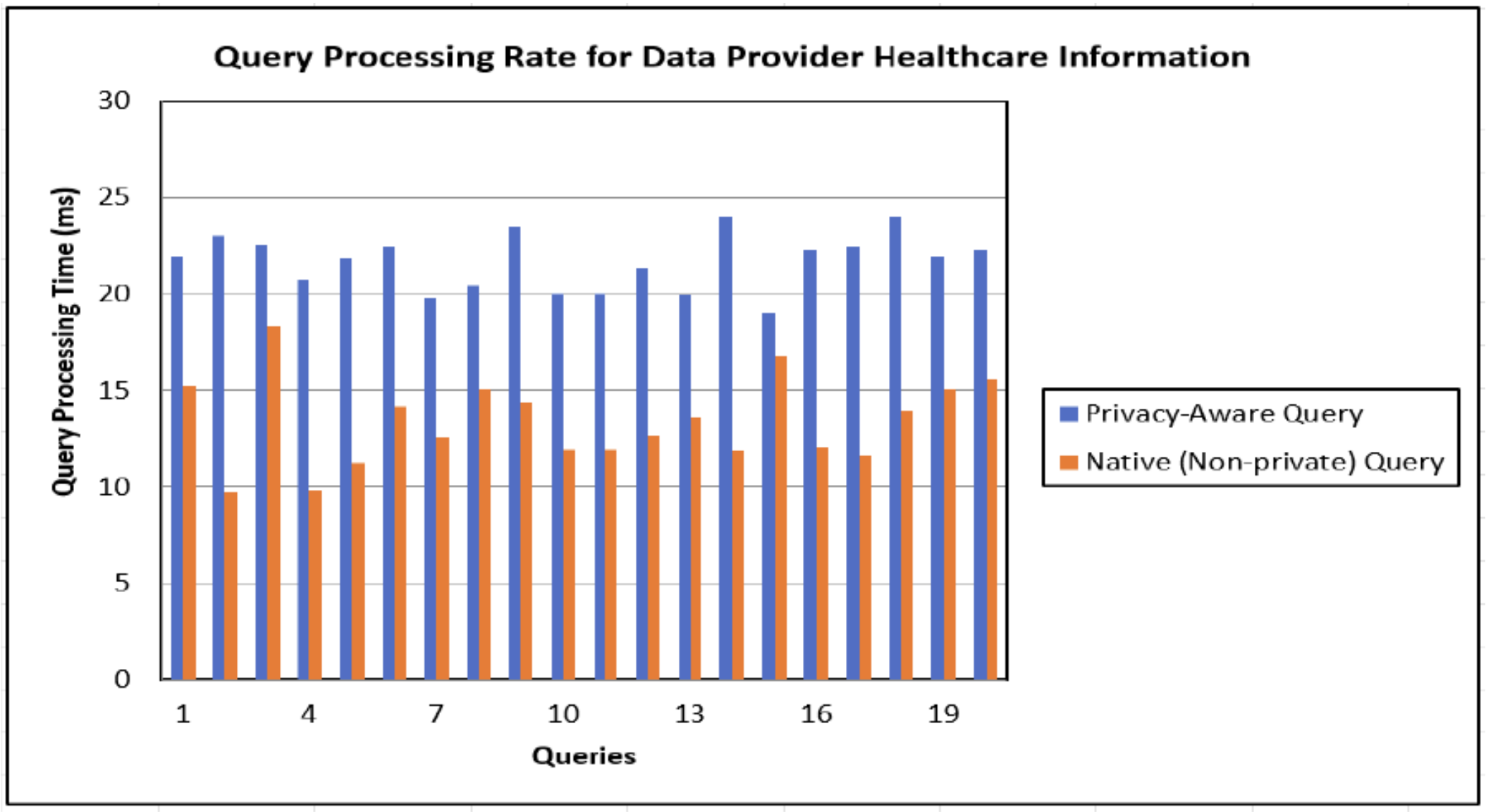

**Figure 12:** Query Processing Rate for Data Provider Healthcare Data.

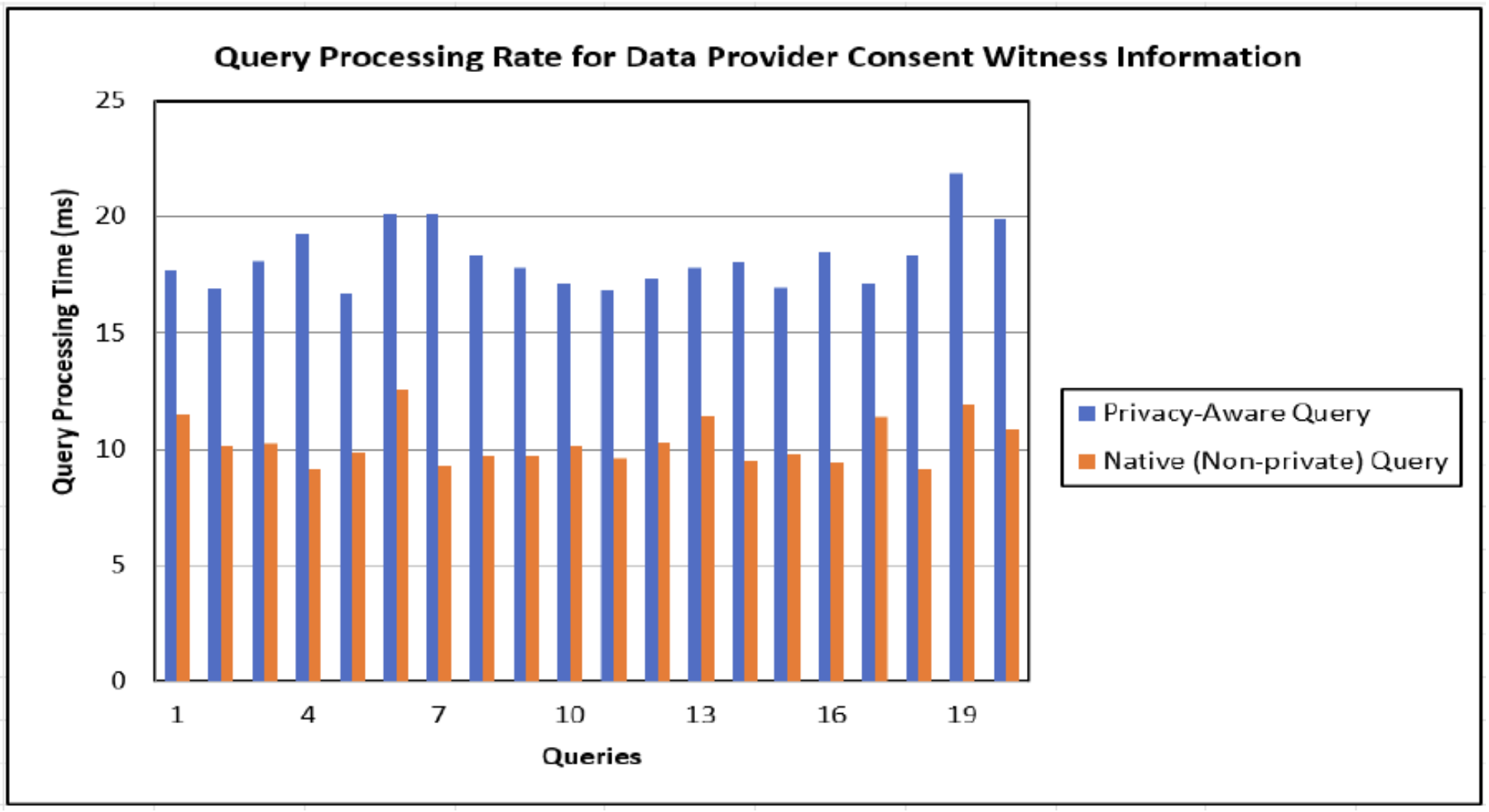

**Figure 13:** Query Processing Rate for Data Provider Consent Witness Data.

architecture offers an efficient platform for processing data provider private information.

### 5.3.3. Query 3 - Data Provider Consent Witness Data

Suppose we want to retrieve data values on data provider health information. The attribute data elements from which the query transaction is performed, include: *Witness Last Name*, *Witness First Name*, *Witness Phone Number*, *Witness Street*, *Witness City*, *Witness Province*, and *Witness Postal Code*, *etc.* The result of query processing rate for data provider consent witness information is displayed in Figure **13**. The privacy-aware queries and native queries are designated with blue and orange colouring, respectively.

**Results Analysis and Discussion:** We observe that query processing rate is in the same range of query response time as the preceding queries. Similar to preceding queries (*i.e.*, *Query 1* and *Query 2*), there is an increase in query processing time on privacy-aware queries in comparison to native (non-private) queries. From the query results, we note that the highest and lowest query processing time for native queries are 12.57 ms (milli seconds) and 9.15 ms (milli seconds), respectively. Moreover, an average query processing time of 10.28 ms is attained. In terms of privacy-aware query processing, we attain the highest and lowest query processing time of 21.85 ms and 16.71 ms, respectively. We also record an average query processing time of 18.24 ms.

We observe privacy overhead cost of 43.65% for privacy-aware query processing. This is attributed to data retrieval that involves greater number of attribute data elements in comparison to queries *Query 1* and *Query 2*. Moreover, the privacy-aware query parsing and analyzes involving these attribute data contribute to the increase in privacy query overhead cost. An analysis on the data provider privacy preferences indicate parameters of data information privacy with instance data representation of: *'Level-2: Internal Use'*, granularity privacy with instance data representation of: *'Existential: Information released to indicate the existence of data in repository, not actual data values'*, visibility privacy with instance data representation of: *'House: Data accessible to everyone who collects, access, and utilize the data'*, and purpose privacy with instance data representation of: *'Reuse-Selected: Data used for primary purpose for which data'*. We determine that based on the query response times attained from both privacy-aware and native query processing, the proposed privacy model offers an efficient platform for data provider privacy-aware query processing.

### 5.3.4. Average Query Processing and Privacy Overhead Cost Analysis

We compute average query processing rate for all three queries and analyse the query response time for each query form. We note that there is a considerable increase in the query response time for privacy-aware queries in relation to native queries. This is mainly attributed to privacy overhead cost from privacy-aware query parsing, analyzes, and execution.

Table **1** illustrates the average query response time and their respective privacy overhead cost. Here, we note that queries, *Query 1* (Data Provider Demographic Data), *Query 2* (Data Provider Health Data), and *Query*

**Table 1: Summary of Average Query Response Time and Privacy Overhead Cost**

| Queries | Average Query Response Time (ms) | | |
|---|---|---|---|
| | Native (Non-private) Query | Privacy-Aware Query | Privacy Overhead Cost |
| Query 1: Data Provider Demographic Data | 13.37 | 22.99 | 9.62 (41.85%) |
| Query 2: Data Provider Healthcare Data | 13.38 | 21.67 | 8.29 (38.26%) |
| Query 3: Data Provider Consent Witness Data | 10.28 | 18.24 | 7.96 (43.65%) |
| All Queries | 12.34 | 20.96 | 8.62 (41.25%) |

*3* (Data Provider Consent Witness Data) deliver privacy overhead query response time of 9.62 ms, 8.29 ms, and 7.96 ms, respectively. Moreover, an average privacy overhead cost of 8.62 ms (milli seconds) is realized for all queries processed. These privacy overhead query response times translate into 41.85%, 38.26%, and 43.65% for queries: *Query 1*, *Query 2*, and *Query 3*, respectively. We note an average of 41.25% privacy overhead query cost for all queries processed. We illustrate the average query processing time and privacy overhead cost for all queries in Figure **14**.

In general, we note that the average privacy overhead query cost is relatively very low. This indicates an effective privacy-aware query processing platform from the privacy infrastructure; and infers an optimum run of privacy-aware queries in comparison to native queries. A review of attained privacy overhead query cost indicates that the proposed privacy infrastructure, and resultant privacy overhead query cost does not excessively or negatively impact on overall query transaction processing.

Moreover, it will be noted that the privacy overhead cost attained for the queries processed are related to the expected operation conditions of the computing system. There will be variations in the expected privacy overhead cost for queries processed in different operating conditions, particularly in cases where computing resources vary (in either higher memory size, number of processors, processor speed, and distributed multi-node blockchain architecture, amongst others), and database size and query volumes vary.

Hence, we establish that the proposed privacy model offers an efficient, reliable, and robust framework for privacy-aware query processing on data provider private and transaction data stored in the blockchains and relational database, respectively.

## 6. COMPARISON TO OTHER METHODOLOGY APPROACHES

There have been some recent significant studies in the area of using blockchain platforms to enforce privacy on data providers' private and sensitive data. These approaches present important contributions with regards to various data protection measures against unauthorized data access to data provider's private data. In comparison to our methodology, our approach addresses the privacy-preservation problem from important concepts of data element coupling, privacy-preservation modelling using privacy ontologies, and data store coupling. We comparatively explain how our methodology approach offers efficient and better measures for data provider's private data protection and identity preservation.

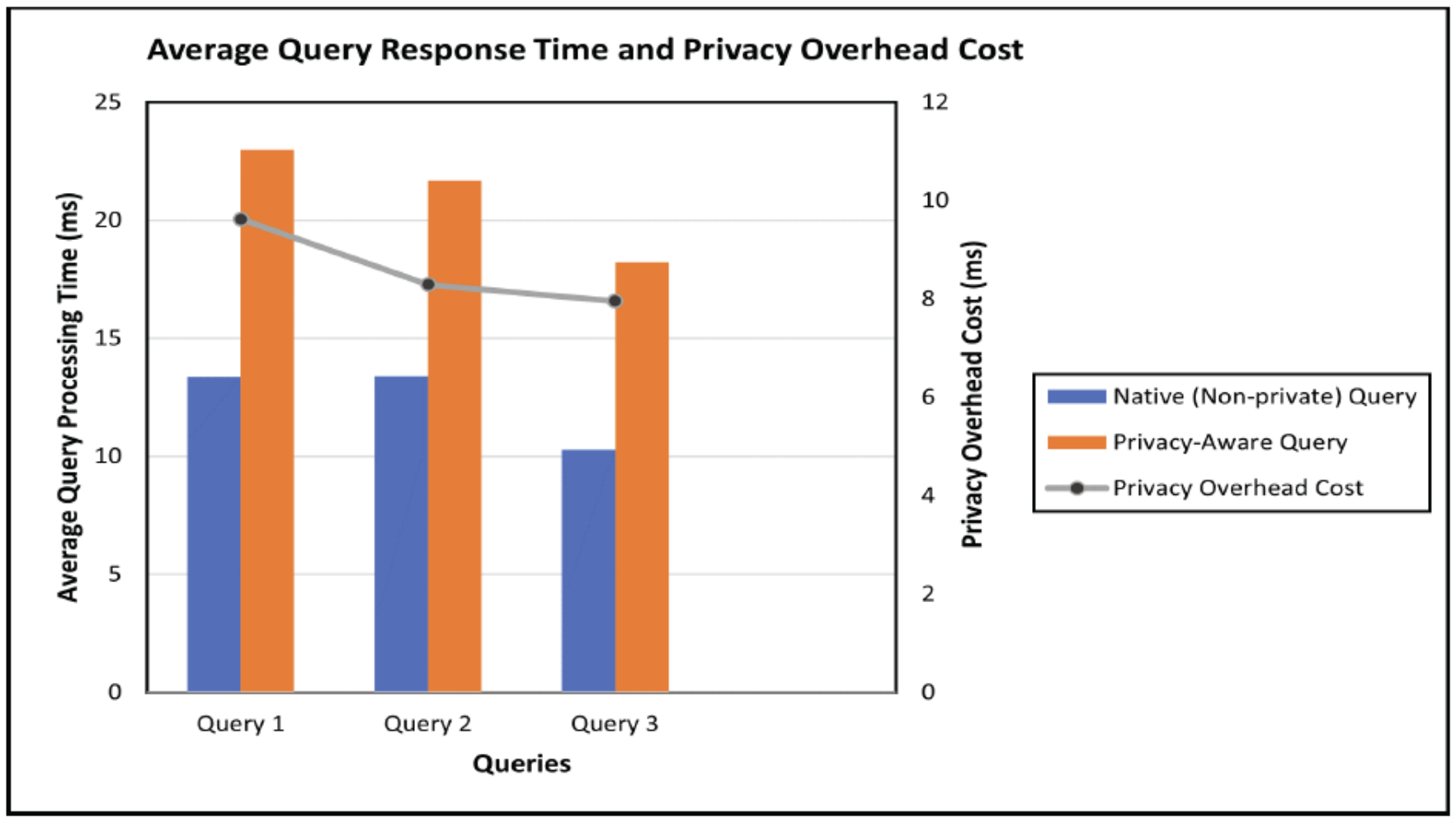

**Figure 14:** Average Query Processing Time and Privacy Overhead Cost.

### 6.1. Daidone *et al.* [27]: Blockchain-based Privacy Enforcement in IoT domain

Daidone *et al.* [27] propose a blockchain-based privacy enforcement architecture where users can define how their data are collected and managed, and ascertain how these data are used without relying on a centralized manager. Their methodology adopts a blockchain to perform privacy preferences' compliance checks in a decentralized fashion on IoT devices; whiles ensuring the correctness of the process through smart contracts. The overall system is based on a privacy model designed for IoT systems and which has been outlined in Sagirlar *et al.* [43]. This privacy model supports standard privacy-related elements (*e.g.*, purpose). Moreover, the privacy model supports set of features to allow data owners to limit how and which data can be retrieved during IoT analytic processes. The privacy model allows the automatic generation of privacy preferences for newly derived data (*e.g.,* information resulting from data fusion). The research considers the privacy model to focus mainly on the traditional privacy preference components (*e.g.*, purpose, retention time).

The system architecture identifies the registration of both data owner (*i.e.*, user) and consumer devices on the blockchain. The smart environment at the data owner's device and consumer servers constitute the IoT network. The IoT network senses data from the user environment and sends these data to the consumer servers. For each of the data owner's IoT network and consumer IoT network, there are attached gateway blockchain nodes. Data collected by the gateway nodes are complemented with metadata that encodes the data owner's privacy preferences. Consumers register their privacy policies into the blockchain. For each IoT device, manufacturers specify system-defined privacy preferences. On the other hand, the data owner is free to add further restrictions to system-defined privacy preferences; by specifying owner-defined privacy preferences. The system architecture relies on smart contracts for the enforcement of privacy preferences. The gateway nodes (which contains the smart contracts) verifies whether the consumer policy satisfies the constraints specified by the data owner in his/her privacy preferences. The proposed architecture leverages on a permissioned blockchain that is configured such that only given stakeholders are authorized to join the privacy compliance check. The system adopts Hyperledger Fabric blockchain.

### 6.2. Fernandez *et al.* [28]: Privacy-Preserving Architecture for Cloud-IoT

Fernandez *et al.* [28] propose a cloud-IoT architecture, called *Data Bank*, that protects users' sensitive data by allowing them to control which kind of data is transmitted by their devices. The system architecture provides users (*i.e.,* data owners) with mechanisms to specify data-collection policies at device level and data-sharing policies at cloud level. Both cloud and local data repositories to allow users to keep their private data safe. Before data is transferred to the cloud repository, it will be temporarily stored in the local *Data Pocket*. The *Data Pocket*, which is under the user's control, consists of memory (*e.g.,* disk storage) and a microprocessor to keep pre-defined data collection policy and filter users' data before uploading to the cloud repository.

The system architecture contains a privacy-utility mechanism. This mechanism aims at finding the right balance between benefits gained and privacy lost when data is provided to external services. It recommends services to users based on users' pre-defined privacy criteria. Users can view and customise their privacy policy via the interface provided. The system offers enforcement of access control policies to restrict access to users' data by third parties. The system architecture consists of a four-tier architecture for cloud-IoT platforms: *application*, *cloud layer*, *data pocket*, and *sensors*. The application layer is the topmost layer and consists of a *user interface*. Moreover, there is a *data-sharing interface/manager*, which is considered as the application program interface (API) and controls the data visible to the services. The cloud layer contains five main components: (a) *access control enforcement module*, (b) *auditing module*, (c) *repository*, (d) *privacy-utility mechanism*, and (e) *Privacy policy*. The *data pocket* layer runs in a local computing device (*e.g.*, IoT hub) and deals with the data-collection policies. In summary, the system provides a privacy preservation architecture where the user has control of the data collection and data sharing policies enforced, and the user can update the policies at any time.

### 6.3. Griggs *et al.* [29]: Healthcare Blockchain System using Smart Contracts for Secure Automated Remote Patient Monitoring

Griggs *et al.* [29] propose a healthcare blockchain system for secure automated remote patient monitoring. In their approach, the research outlines using blockchain-based smart contracts to facilitate secure analysis and management of medical sensors. The overall system architecture uses a private blockchain based on Ethereum protocol. The system architecture approach involves using sensors to communicate with a smart device that calls smart contracts, and writes records of all events on the blockchain.

The system activity flow is described as follows. A data provider (*i.e.,* patient) is equipped with various medical devices, such as blood pressure monitor, and is remotely monitored by a service provider (*i.e.,* doctor). The raw data is sent to a master *"smart device,"* typically a smartphone or tablet, for aggregation and formatting by the application. Once complete, the formatted information is sent to the relevant smart contract. In the Ethereum blockchain protocol, the source for information fed to the smart contracts is known as the *"Oracle"*. The *Oracle* in the smart device (master device) communicates directly to the smart contracts. The smart contract will then evaluate the provided data and issue alerts to both the patient and healthcare provider, as well as automated treatment instructions for the actuator nodes, if desired. Through this approach, no confidential medical information is stored on the blockchain or in the smart contracts because government privacy compliance specifications. The actual data transaction measurements are forwarded to a designated Electronic Healthcare Records (EHR) storage database. Moreover, the trace of new transactions are added to the blockchain.

**Table 2: Qualitative Analysis of Proposed Methodology and Other Approaches**

| **Methodology Approach & Criteria** | **(1) Proposed Methodology** | **(2) Daidone *et al.* [27]** | **(3) Fernandez *et al.* [28]** | **(4) Griggs *et al.* [29]** |
|---|---|---|---|---|
| (1) Data Storage Platform | Relational database | IoT device platform | Cloud-IoT platform | Electronic Healthcare Records (EHR) storage database |
| (2) Blockchain Platform Technology | Multichain (permissioned or private blockchain platform accessible by data providers and data collectors/accessors) | Hyperledger Fabric (permissioned or private blockchain platform accessible by data owner and consumer devices) | No defined blockchain platform. Uses a cloud repository for *privacy utility mechanism* | Ethereum protocol (private and consortium-led blockchain platform) |
| (3) Privacy Model | Adopts a formal (dynamic) contextualized privacy ontology that applies to different application domains and platforms | Adopts privacy model defined in Sagirlar *et al.* [43] (that handles *purpose* and *retention time* privacy), and is designed for IoT devices | No clearly defined privacy model. Adopts a cloud layer of different components (of privacy utility mechanism and privacy policy reference) for privacy management | No clearly defined privacy model. Uses a smart contract embedded in a blockchain network node to enforce a privacy model |
| (4) Privacy Methodology & Infrastructure | Tightly couple data elements (attribute data, privacy preferences data, and data accessor data) to generate a *privacy tuple*.<br>Tightly couple relational database with blockchain | The architecture involves user IoT smart devices, blockchain platform, and consumer (*i.e.*, service provider) systems servers. Each of the blockchain nodes are embedded with smart contracts.<br>A gateway node processes data and enforces privacy preservation. | System architecture involves cloud and local data repositories and data provider IoT devices. System architecture consists of four-tier platform: application, cloud layer, data pocket, and sensors.<br>Architecture controls user data to cloud services | System architecture involves coupling of master (smart) device, EHR database, and Ethereum protocol (private and consortium-led blockchain platform).<br>Smart contract is embedded in the private blockchain platform |
| (5) Data/Query Processing Approach | Privacy preferences are stored in the blockchain platform. The blockchain platform serves as an access control and query authentication medium. All queries are processed through the blockchains to ensure privacy preservation | Privacy preferences are stored in the data owner blockchain gateway node. The gateway node processes and authenticates data values before data moves to consumer | *Data pocket* repository keeps pre-defined data collection policy and filters users' data before uploading to the cloud repository.<br>Privacy-utility mechanism determines data retrieval based on users' pre-defined privacy criteria | Smart contract stored in the Ethereum blockchain contains user privacy preferences. Smart contract and blockchain control access and prevent changes to users' data stored in the EHR database. |

Hence, blockchain transactions are linked to the EHR in order to provide authentication of the data in the data provider's medical history as a comprehensive record of healthcare. This authentication helps prevent and detect alterations of a data provider's EHR, whether it be on purpose or accidental. The system has a private and consortium-led blockchain which offers the functionality that only authorized viewers can read the blocks and only designated nodes can execute smart contracts and verify new blocks.

### 6.4. Methodology Evaluation Analysis

We determine that the research work in Daidone *et al.* [27] present relevant propositions and recommendations for devising an efficient privacy methodology to protect data provider's private data. Moreover, the work outlines expressive and practical metrics for privacy preservation in the context of IoT domains. Conversely, our proposed methodology approach identifies data and query processing based on privacy preferences compliance, authentication, and authorization from the blockchain. Our work addresses query recall (*visibility* privacy), query precision (*granularity* privacy), query response time, and privacy overhead cost of privacy-aware queries over corresponding native (non-private) queries. Evaluating and assessing the effectiveness of all methodology approaches, we can deduce that based on the uniqueness of each privacy model, privacy methodology and infrastructure, and type of adopted blockchain platform, it is impractical to compare the performance of data and query processing of our methodology approach in relation to the other approaches.

We present a comparative analysis and evaluation of our proposed methodology in relation to other approaches in Table **2**. This tabular analysis summarizes the discussions regarding methodology approaches addressed in the literature, and outlines the key contributions from each of the approaches.

## 7. CONCLUSION

This paper presents a privacy-preserving data platform for data storage and query processing using blockchains. The research addresses the need for a practical approach to protect data provider's private and sensitive data using immutable, tamper-resistant data platforms.

The adopted privacy model offers an effective integration of data repository components to provide a data platform where data provider's private data and transaction data are proficiently collected, processed, and managed by data collectors. The overall approach formulates a contextualized privacy ontology with unique semantics and properties for privacy preservation. We implement tight-coupling of three data elements (namely, attribute meta-data, data provider privacy preferences data, and data accessor data) into a *privacy tuple*; to be stored in the blockchains. Moreover, we implement tight-coupling of relational database and blockchains platforms. The key reason for this approach is to maximize the merits from all data management platforms. Our evaluation and result analysis establish a better implementation of privacy model and infrastructure. The privacy model offers secure, immutable, and efficient privacy-aware query processing platform. Finally, we discuss that our methodology approach and evaluation results attained validate a better approach for managing data provider's private data in comparison to other methodology propositions.

We envision some areas of open issues and future work. In terms of some open issues; first, we address the choice of private blockchain adopted for the privacy model architecture. It must be noted that different (private) blockchain platforms are incorporated with distinct functionalities for data storage and retrieval. The functionalities deliver merits that tend to offer improved performance or otherwise negatively affect performance with relational database coupling. For example, in our system architecture we adopt Multichain blockchain platform. This blockchain is designed with a data stream transaction ledger functionality, that functions like relational database. This functionality offers efficient storage structure and indexing of transaction data items in the data streams. Second, it is expected that the blockchain is properly secured and protected from unauthorized access. The only medium of transaction processing should be through API connection from the application point-of-contact user interfaces. Any other interaction aside the user interfaces compromises the purpose of the blockchain in the overall system architecture.

Regarding future work, first, we anticipate the execution of complex relational *join* queries on data provider data. Relational *join* queries will improve the overall query processing experience on the tightly-coupled data repositories. Second, we consider the implementation of the methodology on a distributed blockchain architecture to evaluate results from multi-node operations, network delay, and simultaneous users, amongst others.

## DECLARATION OF THE USE OF AI TOOLS

The author declares that they did not use AI tools in writing this paper. The author retained full responsibility for the study design, scientific judgments, verification of references, execution and validation of experiments, interpretation of results, and final manuscript content. No AI-generated or fabricated experimental data were used.

## CONFLICT OF INTEREST

The author declares no conflict of interest regarding the research, authorship, and publication of this article.